\documentclass[prd,aps,twocolumn,floatfix,superscriptaddress]{revtex4}

\def\rmd{{\rm d}}
\def\p{\partial}

\usepackage{amssymb,graphicx,color}

\begin{document}

\title{Excising a boosted rotating black hole with overlapping grids}

\author{Gioel Calabrese}

\affiliation{School of Mathematics, University of Southampton,
Southampton, SO17 1BJ, UK}

\author{David Neilsen}

\affiliation{Department of Physics and Astronomy, Brigham Young
University, Provo, UT, 84602}

\date{\today}

\begin{abstract}
We use the overlapping grids method to construct a fourth order
accurate discretization of a first order reduction of the Klein-Gordon
scalar field equation on a boosted spinning black hole blackground in
axisymmetry.  This method allows us to use a spherical outer boundary
and excise the singularity from the domain with a spheroidal inner
boundary which is moving with respect to the main grid.  We discuss
the use of higher order accurate energy conserving schemes to handle
the axis of symmetry and compare it with a simpler technique based on
regularity conditions.  We also compare the single grid long term
stability property of this formulation of the wave equation with that
of a different first order reduction.

\end{abstract}

\maketitle

\section{Introduction}

Black hole excision has become an important technique in numerical
relativity.  First proposed by Unruh \cite{Unr}, excision consists in
placing an outflow inner boundary which eliminates the black hole
singularity from the domain.  Sometimes combined with
singularity-avoiding slicings, it is used in all of the current
long-term black hole evolutions.  (For review articles see
Refs.~\cite{L,BS,S,A}, some more recent work
includes~\cite{BTJ,ABDGHHH,ABDHPST}.)
Excision
has proved most successful with black holes at a fixed coordinate
location.  For example, in studies of orbiting compact objects (black
holes and/or neutron stars), long runs have been achieved in
co-rotating frames, where dynamic gauge conditions attempt to keep the
black holes at fixed locations.  For more general orbits, and for
simulations over many orbital time scales, it is anticipated that the
ability to move the black holes on the computational grid through a
stable excision algorithm will be crucial.

Excision methods for moving black holes typically require the
extrapolation of data onto the trailing edge of the black
hole~\cite{BBHGCA,Brandt,YoBauSha,ShoSmiSpeLagSchFis,SpeSmiKelLagSho}.
We recently reported on a new excision method using simultaneous,
multiple coordinate patches, and implemented these numerically using
overlapping grids \cite{CN}.  In this method each boundary, whether an
outer boundary or an excision boundary, moving or static with respect
to a main coordinate system, is fixed to at least one coordinate
system.  The different coordinate patches overlap just like the charts
of an atlas.  Information is communicated from one grid to the other
using only interpolation, without any decomposition into ingoing and
outgoing variables.  This excision method has a number of advantages,
including: (1) the possibility of choosing coordinate patches which
conform to each individual boundary, giving smooth numerical
boundaries and thus simplifying the boundary treatment; (2) the
ability to adapt the coordinates near the black hole to the horizon
geometry, allowing the excision surface to be placed relatively far
from the singularity, and thereby ``excising the excisable''; (3) the
use of simple data structures and the implementation of standard
methods for parallelization and adaptive mesh refinement due to the
fact that individual grids are logically Cartesian.

In our previous work we demonstrated this excision method by evolving
a Klein--Gordon massless scalar field on a boosted Schwarzschild
background.  Numerical tests showed that the scheme was stable and
second order convergent for very large boost parameters ($v=0.98c$).
Long-term convergence test of spherical waves in Minkowski space also
showed that the interfaces between the overlapping grids did not introduce
growing numerical errors.  Moreover, our tests indicated that the
overlapping grids technique is robust, in the sense that stability was
not dependent on some of the fine details of the implementation.  For
example, stability was independent of the interpolation method, the
number of grids used, the physical overlap size of the grids (which
was kept fixed while testing convergence), and their relative
resolutions.

Overlapping grids have also been used recently in evolutions of the
full Einstein equations.  Thornburg~\cite{Tho_excision}, for example,
used six spherical coordinate patches to excise a Kerr black hole
($a=0.6$).  The patches are designed such that coordinate lines in one
direction overlay each other, so that the interpolation required is
effectively one dimensional.  Using the BSSN formulation, Thornburg
was able to evolve the Einstein equations for $1500~M$, and it
appeared that the instabilities were related to the outer boundary
treatment.  He also used overlapping patches in an apparent horizon
solver, which is now available in Cactus \cite{Tho_AH}.

More recently, Anderson and Matzner~\cite{And} have performed 
simulations of black
holes with overlapping grids.  They used two stereographic patches to
cover the spherical excision region, and solved the standard $\dot
g$--$\dot K$ ADM equations in a constrained evolution.  They were able
to evolve boosted black holes ($v=0.5c$) across the computational domain,  
and runs of single, stationary black holes ran for more than $700~M$.

Finally, in a related approach,
Reula, Tiglio and Lehner~\cite{ReuTigLeh} use multiple
coordinate patches designed such that all boundary points on
neighboring patches are aligned.  Rather than overlapping, the grids
just touch.  The communication between the touching grids is based on
the Carpenter-Gottlieb-Abarbanel's method \cite{SAT}, which requires
the computation of the characteristic variables.  They successfully
evolved a Klein--Gordon scalar field on a Kerr background spacetime
and are currently working towards fully relativistic evolutions.

In this paper we expand upon our previous work, and choose to study
improvements in the following directions: (1) we generalize the
background spacetime geometry to allow for spinning black holes, and
investigate the effect of high spin ($a=0.99$) on the system's
stability; (2) we upgrade our derivative and interpolation operators
to give global fourth order accuracy; (3) we introduce a spherical
patch for the outer boundary, removing all boundary corners.

While in \cite{CN} we put strong emphasis on the use of energy
conserving schemes based on difference operators satisfying the
summation by parts rule, here we allow ourselves to explore
discretizations for which a stability proof based on the discrete
energy method is not immediately available. At times, such
discretizations can be much simpler, particularly in the higher order
accurate case.

The paper is organized as follows: In Sec.~\ref{Sec:Overlapping} we
outline the main ingredients of the overlapping grids method.
Sec.~\ref{Sec:wave_equation} is a generalization of Sec.~III.B of
\cite{CN} to the spinning case with the inclusion of a potential in
the wave equation.  The various coordinate systems used and the
regularization of the equations on the axis are discussed in
Sec.~\ref{Sec:Kerr}.  Before describing in
Sec.~\ref{Sec:Discretization} the details of the discretization, in
Sec.~\ref{Sec:LC} we produce group velocity diagrams which illustrate
how information propagates through the domain.  A high resolution
convergence test using the forcing solution method is described in
Sec.~\ref{Sec:NumExp}.  This paper contains three appendices.

\section{Overlapping grids}
\label{Sec:Overlapping}

\begin{figure}[ht]
\begin{center}
\includegraphics*[height=12cm]{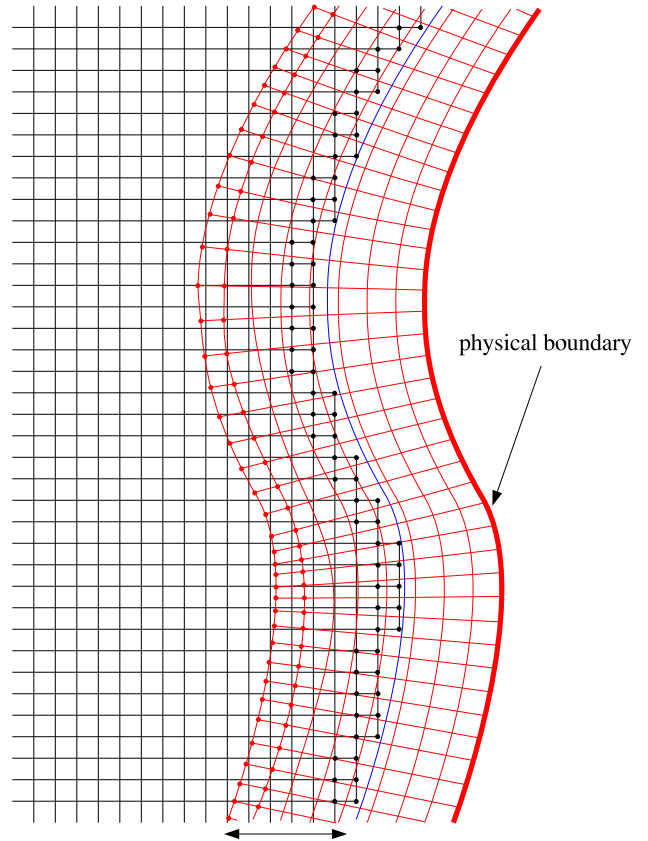}
\caption{This illustrative example consists of two overlapping grids.
The one on the left is the main grid.  The other one is adapted to the
physical boundary of the problem, which may be moving with respect to
the main grid.  The gridpoints marked with a solid dot are updated via
interpolation.  The overlapping grids algorithm is described in the
body of the paper.}
\label{Fig:overlapgeneral}
\end{center}
\end{figure}

When solving hyperbolic initial-boundary value problems numerically,
it is often difficult, if not impossible, to accurately represent the
entire domain, boundary included, with a single grid.  For example, at
least two coordinate patches are required to cover the entire surface
of a 3-sphere without coordinate singularities.  When solving for the
spacetime representing a binary black hole collision, one may want to
use a spherical domain with the outer boundary placed sufficiently far
away, in which the black hole singularities are excised.  To represent
the interior of a sphere with two (or one after the merger) smaller
spheres removed, more than one coordinate system is needed, specially
if one wants the coordinates to be adapted to the boundaries of the
problem.

The overlapping grids methods provides a simple and flexible solution
to such problems.  Our scheme is based on that described in Starius
\cite{Sta} and Ref.~\cite{GKO-Book}.  To demonstrate how the algorithm
works we consider the first order linear hyperbolic system
\begin{equation}
\p_t u = P(t,x,\p_x) u\,,
\label{Eq:u_tPu}
\end{equation}
where $u$ is a vector valued function representing tensor field
components, in a domain $\Omega(t)$ with smooth boundary
$\p\Omega(t)$.  The problem includes appropriate initial and boundary
data.  In general, system (\ref{Eq:u_tPu}) is specified in a
coordinate system $\{t,\vec{x}\}$ which is not and cannot be adapted
to the boundary of the problem.  We will refer to this boundary as the
physical boundary, and we allow for it to be moving with respect to
the main coordinate system.  We introduce a coordinate system
$\{t',\vec{x}'\}$ adapted to $\p\Omega(t)$, in the sense that the
boundary surface can be represented by $x'^i = \rm{const.}$, for some
$i$.  The equations in this coordinate system will take a different
form
\begin{equation}
\p_{t'} u'= P'(t',x',\p_{x'}) u'\,,
\label{Eq:u_tPu2}
\end{equation}
where the components of $u'$ are related to the components of $u$ via
tensor transformation laws.  The initial and boundary data are
also transformed.  To maintain simultaneity of patches we
restrict the coordinate transformation by demanding that $t'=t$.

The problem is discretized by introducing a grid for each coordinate
system.  Any grid point of the main grid which lies beyond a $x'^i =
{\rm const.}$ line (or surface) are dropped, as illustrated in
Fig.~\ref{Fig:overlapgeneral}.  The two grids overlap and the physical
overlap size is kept constant when performing grid refinement tests.

The numerical computation of the right hand side of (\ref{Eq:u_tPu})
at a particular gridpoint $\vec{x}_{ij}$ requires information from a
number of gridpoints in each coordinate direction.  With a fourth
order accurate centered difference operator, for example, one needs two
gridpoints in each direction.  Gridpoints of the main grid, at which
the right hand side cannot be evaluated, are updated via interpolation
from the other grid.  All components of $u'$ are interpolated onto the
main grid and the tensor law transformation is used to evaluate the
components of $u$.  A similar procedure is done at the non-physical
boundary of the second grid.  Points that require interpolation are
marked with a solid dot in the figure.  The numerical treatment of the
physical boundary is as that of a single grid boundary.

In our code we use $n$-th order Lagrange interpolation, which in two
dimensions is given by
\begin{equation}
f_{\rm Int}(x,y) = \sum_{i=1}^n \sum_{j=1}^n \prod_{l=1 \atop l\neq i}^n
\frac{x-x_l}{x_i-x_l} \prod_{k=1\atop k\neq j}^n
\frac{y-y_k}{y_j-y_k}f(x_i,y_j) \label{Eq:nLagrange}
\end{equation}
If $f$ is sufficiently smooth,
the interpolating function is a $n$-th order approximation of
$f(x,y)$. 

Note that if the physical boundary moves with respect to the main
grid, gridpoints may have to be dropped from or added to the main grid
and the set of points which require interpolation needs continuous
updating.  This is done at the end of each full time step of the time
integrator of choice (e.g.~fourth order Runge-Kutta).  When grid
points are added to the main grid, the grid adapted to the
boundary is able to provide accurate data for these points.

The overlapping grids method requires artificial dissipation for
stability \cite{OlsPet}.  We use sixth order dissipation, see
Eq.~(\ref{Eq:KOdissip}), which has a seven point stencil in each
direction.  This means that in our code we actually interpolate three,
rather than two gridpoints.
\begin{figure}[t]
\begin{center}
\includegraphics*[height=16cm]{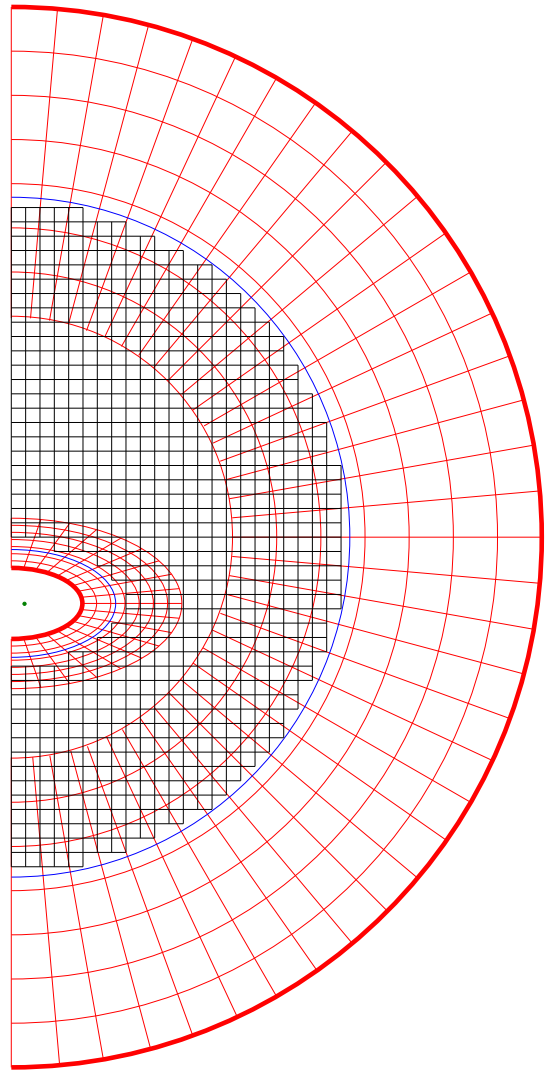}
\caption{We use a main cylindrical grid and two spherical grids
  adapted to the inner and outer boundaries.  The irregular shape of
  the main grid is the result of having dropped gridpoints which lie
  beyond imaginary lines on the spherical grids.  The dot represents
  the ring singularity of the black hole.}
\label{Fig:Overlapping}
\end{center}
\end{figure}

Here we consider the case in which a scalar field propagates on a
boosted spinning black hole background.  Two boundaries are
introduced: an inner and an outer boundary.  The first one represents
the excision surface and is purely outflow.  It requires no boundary
data.  The second one is introduced for computational reasons.  We
need to have a bounded spatial domain because of limited computational
resources.  To handle the two boundaries we introduce two additional
coordinate patches.  One patch is adapted to the outer boundary and
one patch is co-moving with the black hole, and fixed to the inner
excision boundary.  We choose cylindrical coordinates for the main
coordinate patch.  The black hole is boosted with velocity $\beta$
along the axis of symmetry with respect to this coordinate system.
Spheroidal coordinates are used on the second patch, such that the
location of the event horizon is at a constant coordinate value, and
spherical coordinates on the outer patch.  We require that all data in
all three coordinate systems be simultaneous.  By adapting these
coordinates to the black hole horizon, we may excise the spherical
grid at the event horizon for all values of the boost parameter.  This
allows for an efficient use of the excision technique, as we can
excise the excisable.  The layout of the grids is illustrated in
Fig.~\ref{Fig:Overlapping}.  For the fourth order accurate case, which
is the minimum order that we require, Eq.~(\ref{Eq:nLagrange}) 
on an uniform grid takes the form
\begin{eqnarray}
&&f_{\rm Int}(x_i+a \Delta x, y_j + b \Delta y) =\\
&&\sum_{p=i-1}^{i+2} \sum_{q=j-1}^{j+2} \prod_{l=i-1\atop l \neq p}^{i+2}
  \frac{a+i-l}{p-l} \prod_{k=j-1\atop k \neq q}^{j+2}
  \frac{b+j-k}{q-k} f_{pq} \nonumber
\end{eqnarray}
where $0\le a < 1$ and $0 \le b < 1$.

The boundary treatment and the discretization near the axis of
symmetry are described in Sec.~\ref{Sec:Discretization}.

\section{A first order reduction of the wave equation}
\label{Sec:wave_equation}

In this section we write down the wave equation on a general curved
background and discuss a particular first order reduction.  In
Appendix \ref{App:Pireduction} we consider a different first order
reduction, which has different features, and compare the numerical
stability of the two formulations.  For a definition of strong and
symmetric hyperbolicity we refer the reader to
\cite{GKO-Book,KL-Book}.  The rest of this section is a generalization
of Sec.~III.B of \cite{CN}.

The equation of motion for a scalar field propagating on a curved
background $(M,g)$ is given by the second order wave equation
\begin{equation}
\nabla_{\mu} \nabla^{\mu} \Phi - \frac{dV}{d\Phi}= 0,
\end{equation}
where $\nabla$ denotes the covariant derivative associated with the
Lorentz metric $g$ and $V(\Phi)$ is a potential. 

In terms of the tensor density $\gamma^{\mu\nu} =
\sqrt{-g}g^{\mu\nu}$, where $g = \det (g_{\mu\nu})$, the wave equation
can be written as
\begin{equation}
\partial_{\mu} (\gamma^{\mu\nu} \partial_{\nu} \Phi) -\sqrt{-g}
\frac{dV}{d\Phi}= 0\,.
\label{Eq:wave2}
\end{equation}
If we introduce the auxiliary variables $T= \partial_t \Phi$ and $d_i
= \partial_i \Phi$, we can rewrite Eq.~(\ref{Eq:wave2}) as a first
order system,
\begin{eqnarray}
\partial_t \Phi &=& T\,,
\label{Eq:WEgen1}\\
\partial_t T &=& -\Big(\gamma^{ti}\partial_i T +
\partial_i(\gamma^{it}T)+ \partial_i (\gamma^{ij}d_j) +
\label{Eq:WEgen2}\\
&&+\partial_t \gamma^{tt}T + \partial_t
\gamma^{tj}d_j - \sqrt{-g} \frac{dV}{d\Phi}\Big)/\gamma^{tt},\nonumber\\
\partial_t d_i &=& \partial_i T\,.
\label{Eq:WEgen3}
\end{eqnarray}
An attractive feature of this particular first order reduction is that
the constraint variables propagate trivially, namely $\partial_t C_i =
0$.  This ensures that any solution of
(\ref{Eq:WEgen1}--\ref{Eq:WEgen3}) which satisfies the constraints
initially, will satisfy them at later times, even in the presence of
(static) boundaries.

The characteristic speeds in an arbitrary direction $\vec n$, with
$|\vec n| = 1$, are given by the eigenvalues of the matrix
\begin{equation}
A^n = A^i n_i = \left(
\begin{array}{ccc}
0 & 0 & 0\\
0& -2\gamma^{tn}/\gamma^{tt} & -
 \gamma^{nj}/\gamma^{tt}\\
0 & n_i & 0
\end{array}
\right).
\end{equation}
These are $s_\pm = (\gamma^{tn} \pm \sqrt{(\gamma^{tn})^2 -
  \gamma^{tt}\gamma^{nn}})/(-\gamma^{tt}) = \beta^n \pm \alpha
\sqrt{h^{nn}}$ and $s_0 = 0$ with multiplicity equal to the spatial
dimension of the problem, where $\alpha$ is the lapse function,
$\beta^i$ the shift vector, and $h_{ij}$ is the induced 3-metric on
the $t=\mbox{const.}$ slices in the Arnowitt-Deser-Misner (ADM)
decomposition (\ref{Eq:ADMmetric}).  Hyperbolicity requires that the
characteristic speeds be real, namely,
$(\gamma^{tn})^2 - \gamma^{tt}\gamma^{nn} = \det(h_{ij}) \,h^{nn} \ge
0$ for any $n$, which will be true as long as the $t=\mbox{const.}$
hypersurfaces are spacelike.

One can verify that
\begin{equation}
H(t,\vec{x}) = \left(
\begin{array}{ccc}
\xi & 0 & 0 \\
0 & -\eta\gamma^{tt} & 0 \\
0 & 0 & \eta \gamma^{ij}
\end{array}
\right)
\label{Eq:symm}
\end{equation}
where $\xi$ and $\eta$ are functions of $t$ and $\vec x$, is the most
general symmetric matrix that satisfies $HA^i = (HA^i)^T$, $i=1,2,3$.
When positive definite, which will be the case if and only if
$\partial_t$ is timelike and $\xi>0$, $\eta>0$, it represents the most
general symmetrizer of system (\ref{Eq:WEgen1}--\ref{Eq:WEgen3}).
Using the fact that $\gamma^{tt} < 0$ (because the $t={\rm const.}$
slices are spacelike) and $\gamma^{ij} = \sqrt{-g} g^{ij} = \alpha
\sqrt{\det (h_{ij})}(h^{ij} - \beta^i\beta^j/\alpha^2)$ one can show that the
symmetrizer is positive definite if and only if the vector field
$\p_t$ is timelike.

The symmetrizer can be used to construct an energy and obtain energy
estimates.  The time derivative of
\begin{equation}
E = \int_{\Omega} \left[ \frac{1}{2}(-\gamma^{tt}T^2 +
\gamma^{ij}d_id_j) + \sqrt{-g} V(\Phi)\right] \, d^3 x
\label{Eq:simpleE}
\end{equation}
is given by
\begin{eqnarray}
\frac{d}{dt} E &=& \int_{\partial\Omega} \left( T \gamma^{ti} T + T
\gamma^{ij} d_j \right) n_i \, d^2 \sigma \\
&&+  \frac{1}{2}\int_{\Omega} \left(T
\partial_t \gamma^{tt} T + 2T\partial_t \gamma^{tj} d_j + d_i
\partial_t \gamma^{ij} d_j \right) d^3x \nonumber\\
&& + \int_{\Omega}\partial_t \sqrt{-g} V(\Phi) d^3x\,,\nonumber
\end{eqnarray}
where $n_i$ is the outward pointing unit normal to $\p\Omega$.  If
$V(\Phi)$ is quadratic in $\Phi$, e.g.~$V=\frac{1}{2}m^2\Phi^2$,
$\p_t$ is time-like, and maximally dissipative boundary conditions are
used, we have an energy estimate.  Note that in this case
(\ref{Eq:simpleE}) corresponds to the choice $\xi=m^2/2$ and $\eta =
1/2$ in (\ref{Eq:symm}). If, furthermore, the background admits a
time-like vector field $k$ and we use a coordinate system adapted to
it, $\partial_t = k$, the components of $\gamma^{\mu\nu}$ will be time
independent and we have, ignoring boundary terms, energy conservation.

The integrand of the surface term can be written as
\begin{equation}
 2( T \gamma^{ti} T + T \gamma^{ij} d_j) n_i = \lambda_+
{w^{(+\lambda_+;n)}}^2 - \lambda_- {w^{(-\lambda_-;n)}}^2
\end{equation}
where $\lambda_{\pm} = \gamma^n \pm \gamma^{tn}$ and 
\begin{eqnarray}
w^{(\pm \lambda_\pm; n)} &=& \pm \frac{\sqrt{1 \pm
\hat{\gamma}^{tn}}}{\sqrt{2}} T + \frac{1}{\sqrt{2}} \frac{
\hat{\gamma}^{in}d_i}{\sqrt{1 \pm \hat{\gamma}^{tn}}}\
\label{Eq:charvarspm}\\
w^{(0;n)} &=& \gamma_{\perp}^{i}d_i
\label{Eq:charvars0}
\end{eqnarray}
are the orthonormal characteristic variables of $HA^n$.  To simplify
the notation we have introduced the quantities $\gamma^n =
\sqrt{\delta_{\mu\nu} \gamma^{\mu n}\gamma^{\nu n}}$,
$\hat{\gamma}^{\mu n} = \gamma^{\mu n}/\gamma^n$ and
$\gamma^i_{\perp}$.  The latter satisfies $\delta_{ij} \gamma^i_\perp
\gamma^j_\perp = 1$ and $\delta_{ij}\gamma^i_\perp \gamma^{jn} = 0$.
To express the primitive variables in terms of the characteristic
variables we invert Eqs.~(\ref{Eq:charvarspm})--(\ref{Eq:charvars0}),
\begin{eqnarray}
T &=& \frac{\sqrt{1+\hat{\gamma}^{tn}}}{\sqrt{2}} {w^{(+\lambda_+;n)}}
- \frac{\sqrt{ 1 -\hat{\gamma}^{tn}}}{\sqrt{2}} {w^{(-\lambda_-;n)}}
\label{Eq:charvarsT}
 \\ 
d_i &=& \frac{\hat{\gamma}^{in}}{\sqrt{2}} \left(
\frac{{w^{(+\lambda_+;n)}}}{\sqrt{1+\hat{\gamma}^{tn}}} +
\frac{{w^{(-\lambda_-;n)}}}{\sqrt{1-\hat{\gamma}^{tn}}} \right) +
\gamma^i_\perp w^{(0;n)}
\label{Eq:charvarsd_i}
\end{eqnarray}
We use Eqs.~(\ref{Eq:charvarspm})--(\ref{Eq:charvarsd_i}) in the
boundary conditions to prescribe exact data to the incoming
characteristic variable $w^{(+\lambda_+;n)}$.  The variable $\Phi$ is
a zero speed characteristic variable for any direction $n$ and
requires no boundary data.

We also assume axisymmetry, which implies that there exists a
spacelike Killing field $\mbox{\boldmath $\psi$} = \psi^{\mu}
\partial_{\mu}= \partial_{\phi}$.  By adopting coordinate systems
adapted to the Killing field, we have that the metric components are
independent of the $\phi$ coordinate.  Since we are only interested in
axisymmetric solutions of the wave equation, i.e., solutions which do
not depend on $\phi$, the variable $d_{\phi}$, which represents
$\p_{\phi} \Phi$, can be eliminated from the system.

In the next section we define the various coordinate systems used in
our overlapping grids scheme and specialize
Eqs.~(\ref{Eq:WEgen1})--(\ref{Eq:WEgen3}) to these coordinates.

\section{The Kerr metric in Kerr--Schild coordinates}
\label{Sec:Kerr}

We are interested in the case in which the background is given by a
rotating Kerr black hole.  We will use Kerr--Schild coordinates
\cite{KerSch,HE}.  After recalling basic properties of these
coordinates we explicitly compute the coefficients $\gamma^{\mu\nu}$
needed in the evolution system (\ref{Eq:WEgen1})--(\ref{Eq:WEgen3})
and determine the regularized equations on the axis.

The Kerr--Schild metric components are given by
\begin{equation}
g_{\mu\nu} = \eta_{\mu\nu} + 2H\ell_{\mu}\ell_{\nu}
\end{equation}
where 
\begin{eqnarray}
&&\eta_{\mu\nu}\ = {\rm diag} \{ -1,+1,+1,+1\}\,,\\
&&H = \frac{Mr^3}{r^4+a^2z^2}\,,\\
&&\ell_{\mu} = \left( 1, \frac{rx+ay}{r^2+a^2},
\frac{ry-ax}{r^2+a^2}, \frac{z}{r} \right),
\end{eqnarray}
and $r$ is determined implicitly by
\[
\frac{x^2+y^2}{r^2+a^2} + \frac{z^2}{r^2} = 1\,,
\]
or explicitly by
\begin{equation}
r^2 = r^2_{\rm BL}(x,y,z) \equiv \frac{1}{2} \left( \zeta^2 - a^2 \right) + \sqrt{ \frac{1}{4}
\left( \zeta^2 - a^2 \right)^2 + a^2 z^2 }\,,
\end{equation}
where $\zeta^2 = x^2 + y^2 + z^2$.  The inverse metric can be written
as
\begin{equation}
g^{\mu\nu} = \eta^{\mu\nu} - 2H\ell^{\mu}\ell^{\nu}\,,
\end{equation}
where $\ell^{\mu} \equiv \eta^{\mu\nu}\ell_{\nu} =
g^{\mu\nu}\ell_{\nu}$ is a null vector.

The quantities $M$ and $a$ are constants, $M$ representing the mass
and $Ma$ the angular momentum of the black hole as measured from
infinity.  We restrict ourselves to the case $a^2 < M^2$.  We recall
that the event horizon is located at $r = r_+ = M +
(M^2-a^2)^{1/2}$, the Cauchy horizon at $r = r_- = M -
(M^2-a^2)^{1/2}$, and the stationary limit surface, the set of points
in which the Killing field $\kappa = \p/\p t$ becomes null, is given
by $r= M + (M^2 - a^2 \cos^2\theta)^{1/2}$.  Another set of points in
which $\kappa$ is null is given by $r=M-(M^2-a^2\cos^2\theta)^{1/2}$.
The disc $x^2+y^2 \le a^2$, $z=0$ corresponds to $r=0$.  The ring
$x^2+y^2 = a^2$, $z=0$ is a curvature singularity.  For later
convenience we introduce the quantity $\rho_{\rm BL}^2 (r,\theta)
\equiv r^2 + a^2\cos^2\theta$.

\subsection{Boosted cylindrical coordinates}

The main coordinate system is obtained by performing a Lorentz boost,
followed by a transformation to cylindrical coordinates.  Under a
Lorentz boost, i.e., in the new coordinates
\begin{eqnarray}
\bar{t} &=& \gamma(t - \beta z )\\
\bar{x} &=& x\nonumber\\
\bar{y} &=& y\nonumber\\
\bar{z} &=& \gamma (z - \beta t),\nonumber
\end{eqnarray}
where $\gamma = (1-\beta^2)^{-1/2}$, the components of the Kerr--Schild
metric become
\begin{eqnarray*}
g_{\bar\mu \bar\nu} &=& \eta_{\bar\mu \bar\nu} + 2H
\ell_{\bar\mu}\ell_{\bar\nu}\,, \\ \qquad \eta_{\bar\mu\bar\nu} &=&
{\rm diag} \{ -1, +1, +1, +1 \}\,,\\ \qquad \ell_{\bar\mu} &=& \left(
\frac{\hat r}{r}, \frac{r\bar x+ a\bar y}{r^2+a^2}, \frac{r\bar y -a
\bar x}{r^2+a^2}, \frac{\hat z}{r}\right),
\end{eqnarray*}
where $\hat r = \gamma (r+\beta z)$, $\hat z = \gamma (z+ \beta r)$,
$z = \gamma (\bar{z} + \beta \bar{t})$ and $r = r_{\rm BL}(x,y,z) =
r_{\rm BL}(\bar x, \bar y, \gamma(\bar z + \beta \bar t))$.  At time
$\bar{t}$ the singularity is located at $\bar{x}^2 + \bar{y}^2 = a^2$,
$\bar{z} = -\beta \bar{t}$.

We now go to cylindrical coordinates $\{ \bar{t}, \bar{\rho},
\bar{z}, \bar{\phi} \}$, with $\bar{\rho} \cos\bar{\phi} = \bar{x}$
and $\bar{\rho} \sin\bar{\phi} = \bar{y}$, obtaining
\begin{eqnarray*}
\ell_{\bar\mu} &=& \left( \frac{\hat r}{r}, \frac{r\bar
\rho}{r^2+a^2}, \frac{\hat z}{r}, -\frac{a\bar\rho^2}{r^2+a^2}
\right)\,,\\
\eta_{\bar\mu\bar\nu} &=& {\rm diag} \{ -1, +1, +1,
+\bar\rho^2 \}
\end{eqnarray*}
and 
\begin{eqnarray}
\eta^{\bar\mu\bar\nu} &=& {\rm diag} \{ -1, +1, +1,
+\frac{1}{\bar\rho^2} \}\,,\nonumber\\ 
\ell^{\bar\mu} &=& \left( -
\frac{\hat r}{r}, \frac{r\bar \rho}{r^2+a^2}, \frac{\hat z}{r},
-\frac{a}{r^2+a^2} \right),\nonumber\\
\gamma^{\bar\mu \bar\nu} &=&
\bar\rho \left(\eta^{\bar\mu\bar\nu} - 2H
\ell^{\bar\mu}\ell^{\bar\nu}\right)\,,\label{Eq:gamma_cyl_boosted}
\end{eqnarray}
where $z = \gamma(\bar{z} + \beta\bar{t})$ and $r= r_{\rm
BL}(\bar\rho\cos\bar\phi, \bar\rho\sin\bar\phi,\gamma (\bar{z} + \beta
\bar{t}))$.  Now the singularity is located at $\bar \rho = |a|$,
$\bar z = -\beta \bar t$.  Note that in these coordinates the time
derivative of $\sqrt{-g}$ vanishes.

In this coordinate system the symmetrizer associated with this first
order reduction does not lead to a conserved energy
($\partial_{\bar{t}} \gamma^{\bar\mu\bar\nu} \neq 0$), except for
$\beta = 0$.  The region in which the system is symmetric hyperbolic
is determined by the set of points in which $\partial_{\bar t}$ is
timelike, namely
\begin{equation}
-g_{\bar t \bar t} = 1-\frac{2H\hat r^2}{r^2} > 0.
\end{equation}
In the unboosted case the system is symmetric hyperbolic outside the
stationary limit surface, and only strongly hyperbolic inside (see
Fig.~\ref{Fig:lc_cyl_a99_b0}).  For $\beta \neq 0$ the region of lack
of symmetric hyperbolicity increases.  This issue was explored in
greater detail in \cite{CN} for the $a=0$ case.
  
On the axis of symmetry ($\bar{\rho} = 0$) equation (\ref{Eq:WEgen2})
needs to be regularized.  We want to express it in a form which avoids
``$0/0$''.  This can be done by taking the limit $\bar{\rho} \to 0$ in
the equations.  For this purpose it is convenient to introduce the
rescaled quantities
\begin{eqnarray*}
&&\tilde{\gamma}^{\bar{t}\bar{t}} =
\frac{\gamma^{\bar{t}\bar{t}}}{\bar{\rho}},\quad
\tilde{\gamma}^{\bar{t}\bar{\rho}} =
\frac{\gamma^{\bar{t}\bar{\rho}}}{\bar{\rho}^2},\quad
\tilde{\gamma}^{\bar{t}\bar{z}} =
\frac{\gamma^{\bar{t}\bar{z}}}{\bar{\rho}},\quad\\
&&\tilde{\gamma}^{\bar{\rho}\bar{\rho}} =
\frac{\gamma^{\bar{\rho}\bar{\rho}}}{\bar{\rho}},\quad
\tilde{\gamma}^{\bar{\rho}\bar{z}} =
\frac{\gamma^{\bar{\rho}\bar{z}}}{\bar{\rho}^2},\quad
\tilde{\gamma}^{\bar{z}\bar{z}} =
\frac{\gamma^{\bar{z}\bar{z}}}{\bar{\rho}},
\end{eqnarray*}
which have a finite limit for $\bar{\rho} \to 0$ provided that $r \ge
r_0>0$.  The right hand side of (\ref{Eq:WEgen2}) at $\bar\rho = 0$
becomes
\begin{eqnarray*}
\partial_{\bar{t}} \bar{T} &=& \Big( \tilde{\gamma}^{\bar{t}\bar{z}}
\partial_{\bar{z}} \bar{T} + 2 \tilde{\gamma}^{\bar{t} \bar{\rho}}
\bar{T} + \partial_{\bar{z}} (\tilde{\gamma}^{\bar{t} \bar{z}}
\bar{T}) + 2 \tilde{\gamma}^{\bar{\rho} \bar{\rho}}
\partial_{\bar{\rho}} d_{\bar \rho} +\\
&&+ 2 \tilde{\gamma}^{\bar{\rho} \bar{z}}
d_{\bar z} + \partial_{\bar{z}} (\tilde{\gamma}^{\bar{z} \bar{z}}
d_{\bar z}) + \partial_{\bar{t}} \tilde{\gamma}^{\bar{t} \bar{t}} \bar{T}
+ \partial_{\bar{t}} \tilde{\gamma}^{\bar{t} \bar{z}} d_{\bar z}+\\
&&-\frac{dV}{d\Phi}\Big)
/(-\tilde{\gamma}^{\bar{t} \bar{t}})\,.
\end{eqnarray*}

In Sec.~\ref{Sec:LC} we analyze the group velocity of the system and
produce plots which give a graphical representation of how information
propagates throughout the domain.

\subsection{Co-moving spherical coordinate system}

To excise the black hole we introduce a spherical coordinate system
$\{t',r',\theta',\phi'\}$ which is related to the unboosted Cartesian
coordinates $\{t,x,y,z\}$, the coordinates in which the black hole is
at rest, by
\begin{eqnarray}
t' &=& \bar t = \gamma(t - \beta z)\label{Eq:sc1}\\
r' &=& r_{\rm BL}(x,y,z)\nonumber\\
\theta' &=& \tan^{-1} \left(\frac{r_{\rm BL}}{z}\sqrt{\frac{x^2+y^2}{r_{\rm BL}^2+a^2}}\right) = \cos^{-1}
\left(\frac{z}{r_{\rm BL}}\right)\nonumber\\
\phi' &=& \tan^{-1}
\left(\frac{r_{\rm BL}y-ax}{r_{\rm BL}x+ay}\right).\nonumber
\end{eqnarray}
Note that for $\beta = 0$ the time coordinate coincides with the
Kerr--Schild time and not that of the Boyer--Lindquist coordinates
\cite{BoyLin}.  In addition, the azimuthal angle $\phi'$ does not
coincide with the Boyer--Lindquist $\phi_{\rm BL}$, but it is related
to it via
\[
\phi' = \phi_{\rm BL} + a \int (r^2 - 2Mr + a^2)^{-1} \rmd r
\]
 
The inverse coordinate transformation of (\ref{Eq:sc1}) is given by
\begin{eqnarray}
t &=& \frac{t'}{\gamma} + \beta r' \cos\theta'\\
x &=& \sin\theta' \left( r' \cos\phi' - a \sin \phi'\right)\nonumber\\
y &=& \sin\theta' \left( r' \sin\phi' + a \cos \phi'\right)\nonumber\\
z &=& r' \cos\theta'\nonumber
\end{eqnarray}
The coordinates are adapted to the event horizon in the sense that its
location ($r'=r_+$) is time independent and setting $t'=\bar t$ ensures
simultaneity with the main coordinate system.  The components of the
inverse metric can be constructed from 
\begin{widetext}
\begin{eqnarray}
\ell^{\mu'} &=& \left( - \gamma (1+\beta\cos\theta'), 1, 0,0 \right)\\
\eta^{\mu'\nu'} &=& \frac{1}{\rho^2_{\rm BL}} \left(
\begin{array}{cccc}
-\rho^2_{\rm BL} & - \beta\gamma (r'^2+a^2) \cos\theta' & \beta\gamma r'
 \sin\theta' & -a \beta\gamma \cos\theta' \\
\cdot & r'^2 + a^2 & 0 & a \\
\cdot & \cdot & 1 & 0 \\
\cdot & \cdot & \cdot & \sin^{-2} \theta'
\end{array}
\right)
\end{eqnarray}
\end{widetext}
where $\rho^2_{\rm BL} = r'^2 + a^2\cos^2\theta'$.  By definition,
$\gamma^{\mu'\nu'} = \sqrt{-g'}g^{\mu'\nu'}$, where $g^{\mu'\nu'} =
\eta^{\mu'\nu'} - 2H\ell^{\mu'}\ell^{\nu'}$ and
\[
\sqrt{-g'} = \frac{\rho_{\rm BL}^2}{\gamma} \sin\theta'\,.
\]

Note that for $a\neq 0$ there is a region outside the event horizon in
which the system is not symmetric hyperbolic, even when
$\beta=0$.

As we did in the cylindrical case on the axis of symmetry ($\theta'=
0$ or $\theta' = \pi$) we need to regularize the equations.
Introducing the quantities
\begin{eqnarray*}
&&\tilde{\gamma}^{t't'} =
\frac{\gamma^{t't'}}{\sin\theta'},\quad
\tilde{\gamma}^{t'r'} =
\frac{\gamma^{t'r'}}{\sin\theta'},\quad
\tilde{\gamma}^{t'\theta'} =
\frac{\gamma^{t'\theta'}}{\sin^2\theta'},\\
&&\tilde{\gamma}^{r'r'} =
\frac{\gamma^{r'r'}}{\sin\theta'},\quad
\tilde{\gamma}^{\theta'\theta'} =
\frac{\gamma^{\theta'\theta'}}{\sin\theta'},
\end{eqnarray*}
and taking the limit $\theta' \to \theta_0$, where $\theta_0 = 0,
\pi$, the right hand side of (\ref{Eq:WEgen2}) on the axis becomes
\begin{eqnarray}
\partial_{t'} T' &=& \Big( \tilde{\gamma}^{t'r'}
\partial_{r'} T' + \partial_{r'}(\tilde{\gamma}^{t'r'}T') \pm
2 \tilde{\gamma}^{t' \theta'}
T'  \\
&&+ \partial_{r'} (\tilde{\gamma}^{r'r'}
d_{r'})+ 2 \tilde{\gamma}^{\theta'\theta'}
\partial_{\theta'} d_{\theta'}\\
&& - \frac{\rho_{\rm BL}^2}{\gamma} \frac{dV}{d\Phi} \Big)
/(-\tilde{\gamma}^{t' t'}),\nonumber
\end{eqnarray}
where the components of $\tilde{\gamma}^{\mu'\nu'}$ are understood to be
evaluated at $\theta'=0, \pi$.  

Note that in general the characteristic speeds in the rotating case
are higher than those of the non-rotating case.  For example, in the
radial direction at $(r,\theta) = (M,\pi/2)$ they can be as large as
as $4/3$ in the extremal unboosted case $a=M$, $\beta = 0$.  In the
nonrotating case, instead, the same characteristic speeds are bounded
by $1$ outside the event horizon.

\subsection{Outer spherical coordinate system}

In order to have a smooth outer boundary we introduce a spherical
coordinate system $\{\bar{t}, \bar{r}, \bar{\theta}, \bar{\phi}\}$,
which is related to the boosted cylindrical coordinate system through
the time independent transformation
\begin{eqnarray}
\bar{\rho} &=& \bar{r}\sin \bar{\theta} \label{Eq:polartransf1}\\
\bar{z} &=& \bar{r}\cos \bar{\theta} \nonumber
\end{eqnarray}
and its inverse
\begin{eqnarray}
\bar{r} &=& \sqrt{\bar{\rho}^2 + \bar{z}^2}\label{Eq:polartransf2}\\
\bar{\theta} &=& \tan^{-1} \frac{\bar\rho}{\bar z}\nonumber
\end{eqnarray}
which ensure simultaneity.

In this coordinate system we have
\begin{eqnarray*}
\sqrt{-\bar{g}_{\rm s}} &=& \bar r^2 \sin \bar \theta\\
\eta^{\bar{\mu}\bar{\nu}}_{\rm s} &=& {\rm diag} \left\{ -1, +1,
+\frac{1}{\bar r^2}, +\frac{1}{\bar r^2 \sin^2 \bar\theta} \right\}\\
\ell^{\bar\mu}_{\rm s} &=& \left( -\frac{\hat r}{r}, 
\frac{r\bar r}{r^2+a^2}\sin^2\bar\theta + \frac{\hat z}{r} \cos\bar\theta, \right. \\
&& \left. \frac{r}{r^2+a^2}\sin\bar\theta \cos\bar\theta - \frac{\hat
z}{r\bar r}\sin\bar\theta , -\frac{a}{r^2+a^2}\right)
\end{eqnarray*}

As for the co-moving spherical coordinate system, to have the evolution equation
in a form that avoids  ``$0/0$'' it is convenient to introduce
the rescaled quantities
\begin{eqnarray*}
&&\tilde{\gamma}^{\bar t \bar t} =
\frac{\gamma^{\bar t \bar t}}{\sin\bar\theta},\quad
\tilde{\gamma}^{\bar t \bar r} =
\frac{\gamma^{\bar t\bar r}}{\sin\bar\theta},\quad
\tilde{\gamma}^{\bar t\bar\theta} =
\frac{\gamma^{\bar t\bar\theta}}{\sin^2\bar\theta},\\
&&\tilde{\gamma}^{\bar r\bar r} =
\frac{\gamma^{\bar r\bar r}}{\sin\bar\theta},\quad
\tilde{\gamma}^{\bar r\bar \theta} =
\frac{\gamma^{\bar r\bar \theta}}{\sin^2\bar\theta},\quad
\tilde{\gamma}^{\bar\theta\bar\theta} =
\frac{\gamma^{\bar\theta\bar\theta}}{\sin\bar\theta}.
\end{eqnarray*}
We get
\begin{eqnarray*}
\p_{\bar t} \bar T &=& \Big( \tilde\gamma^{\bar t\bar r}\p_{\bar r} \bar T + \p_{\bar r}
(\tilde\gamma^{\bar r\bar t} \bar T) \pm 2 \tilde\gamma^{\bar t \bar
  \theta} \bar T\\
&& + \p_{\bar r} (\tilde\gamma^{\bar r \bar r} d_{\bar r}) + 2
\tilde\gamma^{\bar\theta\bar\theta} \p_{\bar\theta} d_{\bar \theta}  \pm 2 \tilde\gamma^{\bar r \bar\theta}d_{\bar r} \\
&&+ \p_{\bar t}
\tilde\gamma^{\bar t\bar t} \bar T + \p_{\bar t} \tilde\gamma^{\bar
  t\bar r} d_{\bar r} - \bar r^2 \frac{dV}{d\Phi} \Big)/
(-\tilde\gamma^{\bar t\bar t})\,.
\end{eqnarray*}

\subsection{Transformation of variables}

The interpolation procedure which is used to communicate information
between the grids requires the coordinate transformations and the
tensor transformation laws to relate the components of the evolved
fields.  The boosted cylindrical and co-moving spherical coordinate
systems are related by
\begin{eqnarray}
\bar{t} &=& t'\\
\bar{\rho} &=& \sin\theta'\sqrt{r'^2 + a^2}\nonumber\\
\bar{z} &=& \gamma^{-1} r' \cos\theta' - \beta t'\nonumber\\
\bar{\phi} &=& \phi' + \tan^{-1}\left( \frac{a}{r'} \right),\nonumber
\end{eqnarray}
and the inverse transformation
\begin{eqnarray}
t' &=& \bar{t} \\
r' &=& r_{\rm
BL}(\bar\rho\cos\bar\phi,\bar\rho\sin\bar\phi,\gamma(\bar z + \beta
\bar t))\nonumber\\
\theta' &=& \tan^{-1}\left(\frac{\bar{\rho}r_{\rm
BL}}{z\sqrt{r^2_{\rm BL}+a^2}} \right)\nonumber\\
&=& \cos^{-1} \left(
\frac{z}{r_{\rm BL}}\right)\nonumber\\
\phi' &=& \bar{\phi} -
\tan^{-1} \left(\frac{a}{r_{\rm BL}}\right),\nonumber
\end{eqnarray}
where $z = \gamma(\bar z + \beta \bar t)$.

The field $\Phi$ transforms like a scalar and the fields $T$ and $d_i$
transform like components of 1-forms.  In this case we have

\begin{eqnarray}
\bar{T} &=& T' + \gamma \beta \frac{r'^2+a^2}{\rho_{\rm BL}^2}\cos\theta' d_{r'} -
\gamma\beta \frac{r'}{\rho^2_{\rm BL}} \sin\theta' d_{\theta'} \\
d_{\bar \rho} &=& \frac{r'\sqrt{r'^2+a^2}}{\rho^2_{\rm BL}}\sin\theta'
d_{r'} + \frac{\sqrt{r'^2+a^2}}{\rho^2_{\rm BL}} \cos\theta' d_{\theta'}\nonumber\\
d_{\bar z} &=& \gamma \frac{r'^2+a^2}{\rho^2_{\rm BL}}\cos\theta'd_{r'}
-\gamma\frac{r'}{\rho^2_{\rm BL}}\sin\theta' d_{\theta'},\nonumber
\end{eqnarray}
and 
\begin{eqnarray}
T' &=& \bar{T} - \beta d_{\bar z}\\
d_{r'} &=& \frac{r'}{\sqrt{r'^2+a^2}} \sin\theta' d_{\bar \rho} +
\gamma^{-1}\cos\theta' d_{\bar z}\nonumber\\
d_{\theta'} &=& \sqrt{r'^2+a^2} \cos\theta' d_{\bar \rho}
- \gamma^{-1}r' \sin\theta' d_{\bar z},\nonumber
\end{eqnarray}
where $(\bar{T},d_{\bar \rho},d_{\bar z})$ and $(T',d_{r'},d_{\theta'})$ are the fields
on the cylindrical and spherical grids, respectively.

We also need the transformations between the boosted cylindrical and the outer
spherical coordinate systems,
Eqs.~(\ref{Eq:polartransf1})--(\ref{Eq:polartransf2}), and
\begin{eqnarray}
d_{\bar \rho} &=& \sin\bar\theta d_{\bar r} + \frac{\cos\bar\theta}{\bar r}
d_{\bar \theta}\\
d_{\bar z} &=& \cos\bar\theta d_{\bar r} - \frac{\sin\bar\theta}{\bar r}
d_{\bar \theta}
\end{eqnarray}
and 
\begin{eqnarray}
d_{\bar r} &=& \sin\bar\theta d_{\bar \rho} + \cos\bar\theta d_{\bar z}\\
d_{\bar \theta} &=& \bar r \left( \cos\bar\theta d_{\bar \rho} - \sin\bar\theta
d_{\bar z} \right)
\end{eqnarray}

\section{Group velocity diagrams}
\label{Sec:LC}

To have a graphical representation of how information propagates
through the domain it is useful to construct group velocity or light
cones diagrams.  These diagrams are obtained by looking at the group
velocity of the system at selected points of the domain.  We briefly
describe how the procedure works.  More details can be found in
\cite{H,GR,R2}.

Consider a system of quasilinear partial differential equations in two
spatial dimensions
\begin{equation}
\partial_t u = A^i(t,x,u)\partial_i u + B(t,x,u)\,.
\end{equation}
The characteristic speeds, or phase velocities, in the direction $\vec n
= (n_1, n_2)$ are given by the eigenvalues of $A^in_i$.  We focus on
one of them, which we denote by $\lambda(t,\vec x,u,\vec n)$.  The group
velocity is given by its gradient with respect to $\vec n$, i.e.,
\begin{equation}
\vec v_g(t,\vec x,u,\vec n)  = \left( \frac{\partial\lambda}{\partial n_1} ,
\frac{\partial\lambda}{\partial n_2} \right)
\end{equation}

In order to produce light cones plots at a certain time $t_0$, about a
solution $u_0$, we select a number of uniformly spaced points of the
domain, $\vec x_i$, and plot the parametric function
\begin{equation}
\vec x(\tau) = \vec x_i + \alpha \vec v_g (t_0, \vec x_i, u_0, \vec n(\tau))
\end{equation}
where $\vec n(\tau) = (\cos\tau, \sin\tau)$ with $\tau \in [0,2\pi)$
and $\alpha$ is a constant introduced for aesthetic reasons.  It
avoids that the cones either overlap or are too small.

\begin{figure}[ht]
\begin{center}
\includegraphics*[height=12cm]{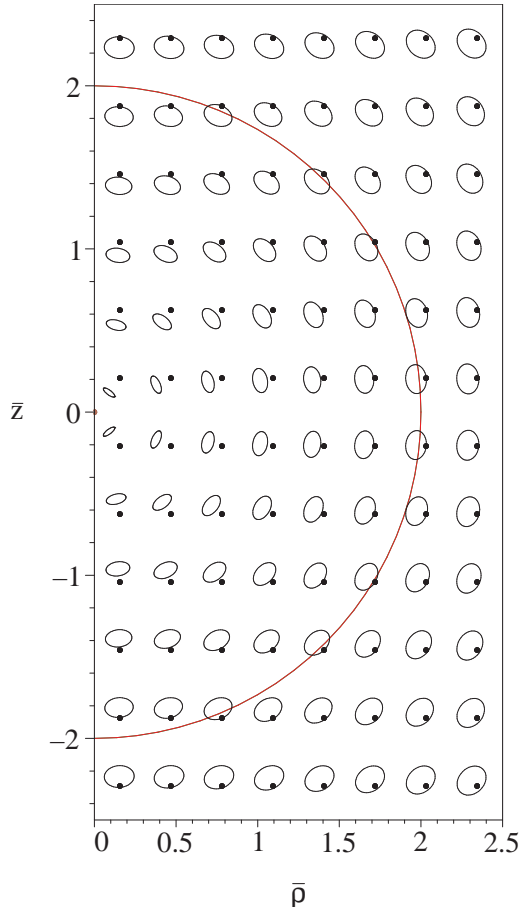}
\caption{The light cone structure in Kerr--Schild cylindrical coordinates for
$\beta =0$, $a=0$, and $M=1$.  The event horizon is located at
  $r=\sqrt{\bar\rho^2 + \bar z^2} = 2M$.}
\label{Fig:lc_cyl_a0_b0}
\end{center}
\end{figure}

\begin{figure}[ht]
\begin{center}
\includegraphics*[height=12cm]{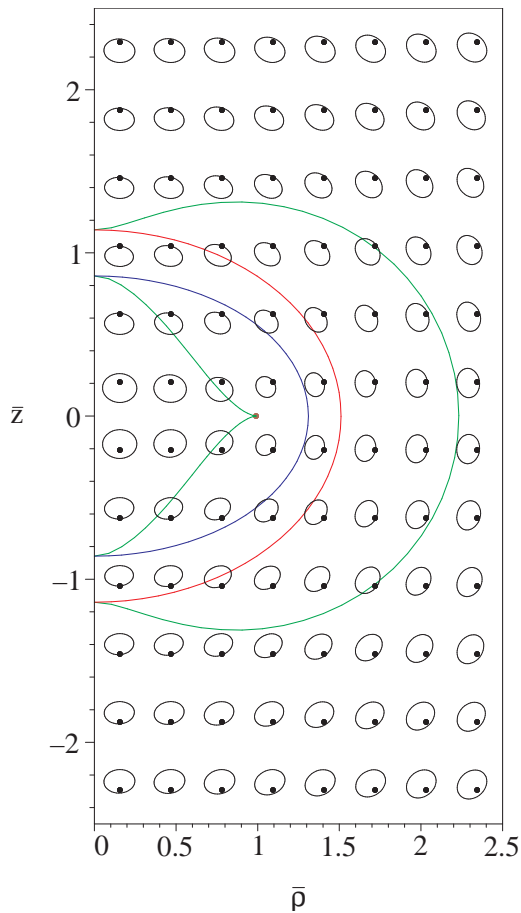}
\caption{The light cone structure in Kerr--Schild cylindrical
coordinates for $\beta =0$, $a=0.99M$, and $M=1$.  The red line
represents the event horizon and the blue line the Cauchy horizon.
The brown dot is the ring singularity.  The outer region between the
event horizon and the green line, the stationary limit surface,
represents the ergoregion.  Between the two green lines the vector
field $\p_t$ is spacelike and, therefore, the system is only strongly
hyperbolic.}
\label{Fig:lc_cyl_a99_b0}
\end{center}
\end{figure}

\begin{figure}[ht]
\begin{center}
\includegraphics*[height=12cm]{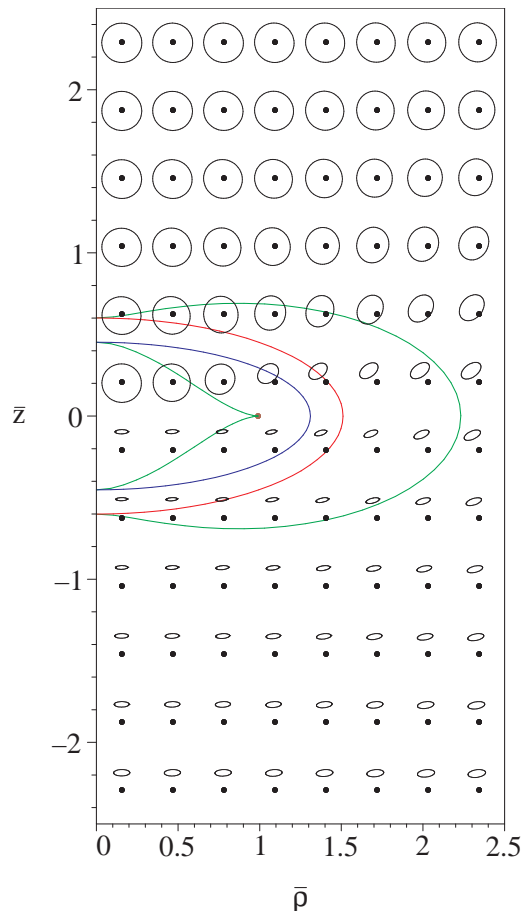}
\caption{The light cone structure in Kerr--Schild cylindrical coordinates for
$\beta =-0.85$, $a=0.99M$, and $M=1$.  The red line
represents the event horizon and the blue line the Cauchy horizon.
The brown dot is the ring singularity, which is moving towards
positive values of $\bar z$.  The outer region between the
event horizon and the green line represents the ergoregion.}
\label{Fig:lc_cyl_a99_b85}
\end{center}
\end{figure}

We produced plots for the linear system
(\ref{Eq:WEgen1})--(\ref{Eq:WEgen3}) in the cylindrical coordinate
system.  Fig.~\ref{Fig:lc_cyl_a0_b0} illustrates the light cone
structure in the non-rotating case, whereas
Fig.~\ref{Fig:lc_cyl_a99_b0} represents the light cones in the rapidly
spinning case, and Fig.~\ref{Fig:lc_cyl_a99_b85} shows the rapidly
spinning, highly boosted case.  It is interesting to see how in
Fig.~\ref{Fig:lc_cyl_a99_b85} the light cones tilt.  While the light
cones that are being approached by the black hole are almost
unperturbed, those behind are tilted in such a way that the
information is forced to follow the black hole.  This ``dragging''
effect is clearly noticeable in our numerical simulations.

\section{Discretization}
\label{Sec:Discretization}

To discretize system (\ref{Eq:WEgen1})--(\ref{Eq:WEgen3}) we replace
the partial derivative $\partial_i$ with the finite difference
operator $D_i$, without expanding derivatives of products.
This leads to the semi-discrete system
\begin{eqnarray}
\partial_t \Phi &=& T\label{Eq:DWEgen1}\\
\partial_t T &=& -\Big(\gamma^{ti}D_i T +
D_i(\gamma^{it}T)+ D_i (\gamma^{ij}d_j) +\label{Eq:DWEgen2}\\
&&+\partial_t \gamma^{tt}T + \partial_t
\gamma^{tj}d_j -\sqrt{-g} \frac{dV}{d\Phi}\Big)/\gamma^{tt},\nonumber\\
\partial_t d_i &=& D_i T\,.
\label{Eq:DWEgen3}
\end{eqnarray}
where we have not explicitly written out the gridfunction indeces.
As was shown in \cite{CN}, this type of discretization leads to
discrete energy conservation provided that $\p_t$ is a time-like
Killing field
\footnote{If $\p_t$ is a Killing field, which is not time-like, we get
  conservation of a non-positive definite quantity.}.

In the interior we use the centered fourth order accurate finite
difference operators
\begin{eqnarray}
D^{(1)}u_{ij} &=& D^{(1)}_0\left(1-\frac{h^2}{6}D^{(1)}_+D^{(1)}_-\right)u_{ij} \label{Eq:D4a}\\
&=& (-u_{i+2,j} + 8u_{i+1,j} - 8u_{i-1,j} + u_{i-2,j})/(12h_1)
\nonumber\\
D^{(2)}u_{ij} &=& D^{(2)}_0\left(1-\frac{h^2}{6}D^{(2)}_+D^{(2)}_-\right)u_{ij} \label{Eq:D4b}\\
&=& (-u_{i,j+2} + 8u_{i,j+1} - 8u_{i,j-1} + u_{i,j-2})/(12h_2)\nonumber
\end{eqnarray}

We now discuss the discretization near the
axis of symmetry and near the physical boundaries.

\subsection{Axis of symmetry}

The difference operators (\ref{Eq:D4a}) and (\ref{Eq:D4b}) have a 5
point stencil.  Gridpoints that are too close to the axis require
special treatment.  If a gridpoint is close to but not on the axis of
symmetry we use Eqs.~(\ref{Eq:DWEgen1})--(\ref{Eq:DWEgen3}) combined
with the regularity conditions to evaluate difference operators in the
direction normal to the axis.  With respect to the coordinate normal
to the axis $\Phi$, $T$, and $d_A$ are even and $d_n$ is odd, where
$n$ is normal to the axis and $A$ is tangent.  If a gridpoint lies on
the axis we use the regularized equations combined with the regularity
conditions.  Specifically, in the boosted cylindrical coordinate case
we use
\begin{eqnarray*}
\partial_{\bar{t}} \bar{T} &=& \Big( \tilde{\gamma}^{\bar{t}\bar{z}}
D^{(\bar{z})} \bar{T} + 2 \tilde{\gamma}^{\bar{t} \bar{\rho}}
\bar{T} + D^{(\bar{z})} (\tilde{\gamma}^{\bar{t} \bar{z}}
\bar{T}) + 2 \tilde{\gamma}^{\bar{\rho} \bar{\rho}}
D_{\rm reg}^{(\bar{\rho})} d_{\bar \rho} + \\
&&2 \tilde{\gamma}^{\bar{\rho} \bar{z}}
d_{\bar z} + D^{(\bar{z})} (\tilde{\gamma}^{\bar{z} \bar{z}}
d_{\bar z}) + \partial_{\bar{t}} \tilde{\gamma}^{\bar{t} \bar{t}} \bar{T}
+ \partial_{\bar{t}} \tilde{\gamma}^{\bar{t} \bar{z}} d_{\bar z}+\\
&& -\frac{dV}{d\Phi}\Big)
/(-\tilde{\gamma}^{\bar{t} \bar{t}}).
\end{eqnarray*}

In the co-moving spherical grid on the axis of symmetry ($\theta =
0$ and $\theta = \pi$) we use the approximation
\begin{eqnarray*}
\partial_{t'} T' &=& \Big( \tilde{\gamma}^{t'r'}
D^{(r')} T' + D^{(r')}(\tilde{\gamma}^{t'r'}T') \pm
2 \tilde{\gamma}^{t' \theta'}
T' + \\
&&D^{(r')} (\tilde{\gamma}^{r'r'}
d_{r'}) + 2 \tilde{\gamma}^{\theta'\theta'}
D_{\rm reg}^{(\theta')} d_{\theta'} +\\
&&-\frac{\rho^2_{\rm BL}}{\gamma}\frac{dV}{d\Phi} \Big)
/(-\tilde{\gamma}^{t' t'}).
\end{eqnarray*}
whereas in the outer spherical grid we use
\begin{eqnarray*}
\p_{\bar t} \bar T &=& \Big( \tilde\gamma^{\bar t\bar r}D^{(\bar r)} \bar T + D^{(\bar r)}
(\tilde\gamma^{\bar r\bar t} \bar T) \pm 2 \tilde\gamma^{\bar t \bar
  \theta} \bar T\\
&& + D^{(\bar r)} (\tilde\gamma^{\bar r \bar r} d_{\bar r}) + 2
\tilde\gamma^{\bar\theta\bar\theta} D^{(\bar\theta)}_{\rm reg} d_{\bar \theta}  \pm 2 \tilde\gamma^{\bar r \bar\theta}d_{\bar r} \\
&&+ \p_{\bar t}
\tilde\gamma^{\bar t\bar t} \bar T + \p_{\bar t} \tilde\gamma^{\bar
  t\bar r} d_{\bar r} - \bar r^2 \frac{dV}{d\Phi} \Big)/
(-\tilde\gamma^{\bar t\bar t})
\end{eqnarray*}
The operator $D^{(n)}_{\rm reg}$ represents the centered fourth order
accurate operator computed using the regularity conditions.  Note that
this discretization is not energy conserving.  This is discussed
further in Appendix \ref{App:Fourth}.  Clearly, the derivative
operator in the direction parallel to the axis needs to be modified
near the physical boundary and boundary data needs to be provided at
the outer boundary.  This is discussed in the next subsection.

\subsection{Boundary conditions}

We always place the excision surface $r'=r'_{\rm min}$ between between
the Cauchy and the event horizon, in which case no boundary conditions
are required.  The fourth order accurate difference operator near the
excision surface in the normal direction to it, however, needs to be
modified.  The modification is given in Appendix \ref{App:43operator}.
A similar modification of the difference operator is required near the
outer boundary of the outer spherical grid, $\bar r = \bar r_{\rm
max}$.  In addition, here we give data to the incoming characteristic
variables in the normal direction, including at those points which lie
on the axis, where the normal is chosen to be parallel to the axis
(see Fig.~4 of \cite{CN}).  We do not couple the ingoing to the
outgoing characteristic variables.  Our numerical implementation is
based on Olsson's method \cite{Ols}.

\subsection{Artificial dissipation}

It is known that overlapping grids require artificial dissipation for
stability~\cite{OlsPet}.  Therefore to the right hand side of the
discretized system we add a term of the form
\begin{equation}
Q_d u_{ij} = \sigma \left( h_1^5 (D^{(1)}_+D^{(1)}_-)^3 +  h_2^5
(D^{(2)}_+D^{(2)}_-)^3\right) u_{ij}\,. \label{Eq:KOdissip}
\end{equation}

We modify this operator near the physical boundaries according to
\begin{eqnarray*}
Q_d u_0 &=& \sigma h^3 D_+^3 u_0\,,\\
Q_d u_1 &=& \sigma h^3 \left( D_+^3 - 3D_-D_+^2\right)u_1\,,\\
Q_d u_2 &=& \sigma h^3 \left( D_+^3 - 3D_-D_+^2 + 3 D_+D_-^2\right)
u_2\,,
\end{eqnarray*}
where we have only kept the relevant gridpoint index.  The
modification near $i=N$ is obtained from the above with the
replacements $i\to N-i$ and $D_{\pm} \to -D_{\mp}$.  This modification
was derived by combining the extrapolation conditions $h^3 D_+^3
u_{-j} = 0$ and $h^3D_-^3u_{N+j}$, $j=1,2,3$ with (\ref{Eq:KOdissip})
whenever gridpoints outside the domain are needed.  The result is then
multiplied by $h$ to ensure that the modification is a third order
correction, which does not lower the global accuracy of the scheme
\cite{Gus}.  Near and on the axis of symmetry dissipation (in the
normal direction) is computed exploiting the regularity conditions of
the fields.  We were not able to modify the dissipative operator such
that a discrete energy estimate holds and without lowering the global
accuracy of the scheme.

\subsection{The discrete energy method}

Unlike for the second order accurate case considered in \cite{CN}, we
are not able to show that our discretization satisfies a discrete
energy estimate, not even on a single grid and with homogeneous
boundary data.  We should stress that the inability to obtain a
discrete energy estimate is by no means a proof of instability.  The
discrete energy method is only a sufficient condition for stability.

In Sec.~\ref{Sec:NumExp} numerical experimentation is used to
determine the stability of our scheme.

\subsection{Choice of time step and dissipation parameter}

\begin{figure}[ht]
\begin{center}
\includegraphics*[height=6cm]{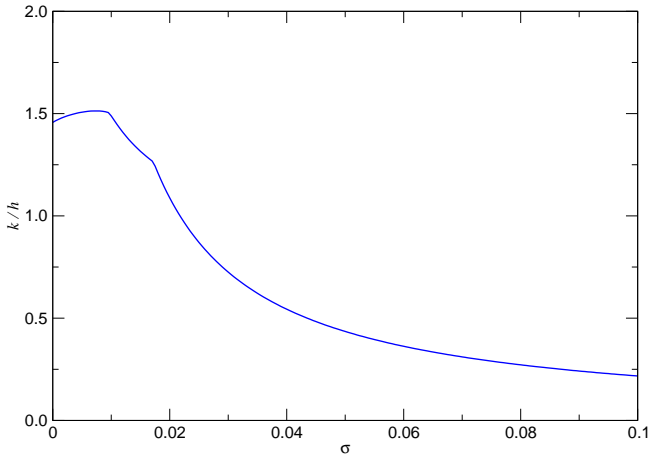}
\caption{The maximum value of the Courant factor for the stability of
the fourth order accurate approximation of the 2D wave equation, $\p_t
u_0 = \p_i u_i$, $\p_t u_i = \p_i u_0$, with artificial dissipation
(\ref{Eq:KOdissip}), integrated with fourth order
Runge-Kutta, as a function of the dissipation parameter
$\sigma$. \label{Fig:courantlimits_2d_4th}}
\end{center}
\end{figure}

We obtain the fully discrete system by integrating the semi-discrete
system of ordinary differential equations with fourth order
Runge-Kutta.  Whenever explicit finite difference schemes are used to
approximate hyperbolic problems, the ratio between the time step size
$k$ and the mesh size $h = \min \{ h_i\}$, the Courant factor, cannot
be greater than a certain value~\cite{CFL}, which is inversely
proportional to the characteristic speeds of the system.  In
Fig.~\ref{Fig:courantlimits_2d_4th} we numerically estimate allowable
values for the Courant factor by examining the 2D wave equation
written in first order form, $\partial_t u_0 = \partial^i u_i$,
$\partial_t u_i = \partial_i u_0$.  We plot the Courant limits for
fourth order Runge-Kutta as a function of the artificial dissipation
parameter, assuming fourth order, centered differencing for the
spatial derivatives (\ref{Eq:D4a})--(\ref{Eq:D4b}) and sixth order
dissipation (\ref{Eq:KOdissip}).  A similar plot for the second order
accurate case can be found in Fig.~7 of \cite{CN}.  From the plot we
see that the value $\sigma \approx 0.0075$, while damping the high
frequency modes, does not force us to reduce the Courant factor.  This
is the value of dissipation parameter that we use in our simulations.
It is about three times smaller than the correspondent value for a
second order accurate approximation ($\sigma \approx 0.02$).

\section{Numerical Experiments}
\label{Sec:NumExp}

In this section we outline the test carried out to verify that our
overlapping grid code is fourth order convergent.  We faced some
unexpected issues when testing for long term stability.  These are
discussed in Appendix \ref{App:Pireduction}.

Let $u$ be the exact solution of the continuum problem at time $t$ and
$v^h$ the solution of the fully discrete approximation at time step
$n$, such that $t=kn$, obtained with a mesh size $h$.  The convergence
rate is computed as
\begin{equation}
Q_h \equiv \log_2 \frac{\| u-v^h \|_h}{\| u-v^{h/2}\|_h}
\label{Eq:convergence}
\end{equation}
where $\| \cdot \|_h$ is a discrete $L_2$-norm.  Neglecting round-off
errors one should expect that $\lim_{h\to 0} Q_h = 4$ for a fourth
order accurate scheme.  To use this equation we need an exact solution
of the continuum problem.  We use the same forcing solution technique
described in \cite{CN}.

Let us rewrite the partial differential equation as $L(u) = 0$ and let
$w$ be an arbitrary function.  If $w$ is inserted into the equation,
in general, it will produce a non vanishing right hand side,
\begin{equation}
L(w) = F\,.
\end{equation}
Clearly, the modified equation $\tilde{L}(u) \equiv L(u) - F = 0$ has
$w$ as an exact solution and the convergence of the code can be tested
using Eq.~(\ref{Eq:convergence}).

\begin{figure}[t]
\begin{center}
\includegraphics*[height=6cm]{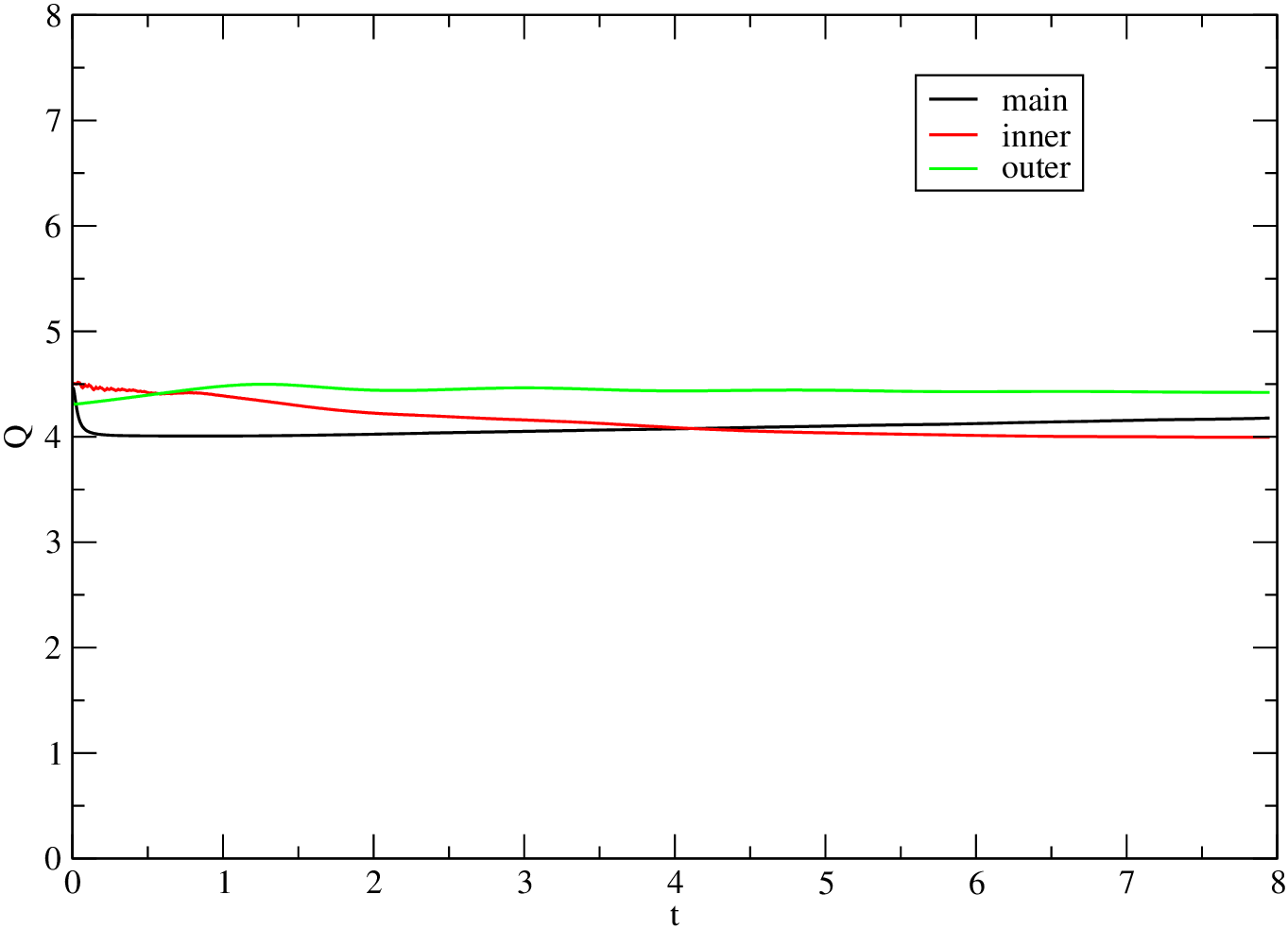}
\caption{The convergence rates $Q_h$ on the three grids obtained using
  the forcing solution $w(t,r,\theta)$ with $n=2$.  The domains and
  coarsest resolutions are: $[0,10]\times[-10,10]$, $160\times 320$
  for the main grid; $[1,3]\times[0,\pi]$, $80\times 240$ for the
  inner spherical grid; $[8,14]\times [0,\pi]$, $240\times 240$ for
  the outer spherical grid.  The Courant factor is set to $0.4$ and
  the dissipation parameter to $\sigma = 0.0075$.  The values of other
  parameters are: $M = 1$, $\beta = -0.85$, $a=0.99$.}
\label{Fig:Qa99_b85}
\includegraphics*[height=6cm]{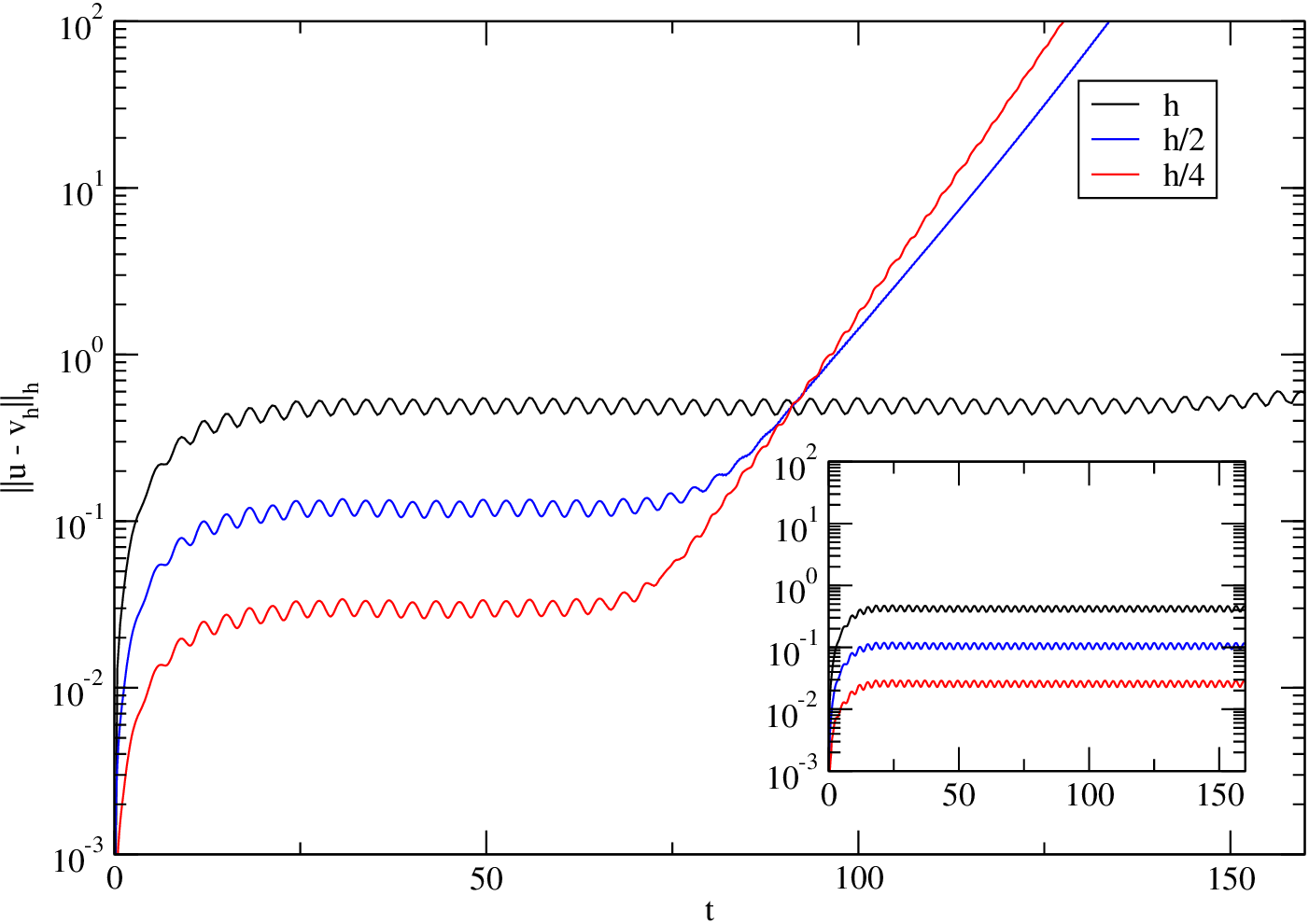}
\caption{This figure illustrates the importance of artificial
  dissipation when using overlapping grids.
  The discrete $L_2$ norm of the error on the main grid at
  three different resolution for $M=a=\beta=0$.  For this test second
  order accurate operators, satisfying summation by parts on each
  grid, were used, but no artificial dissipation was added.  The
  Courant factor is set to $1$.  The domains and coarsest resolutions
  are: $[0,10]\times[-10,10]$, $80\times 160$ for the main grid;
  $[1,3]\times[0,\pi]$, $40\times 120$ for the inner spherical grid;
  $[8,14]\times [0,\pi]$, $120\times 120$ for the outer spherical
  grid.  The lack of stability is evident.  On the other hand, as
  shown in the inset, when artificial dissipation is added to the
  overlapping grids scheme ($\sigma=0.02$) stability is restored.}
\label{Fig:nodiss}
\end{center}
\end{figure}

We chose $w(t,r,\theta) = \sin(t+r) \cos(n\theta)$, where
$\{t,r,\theta\}$ are the spheroidal coordinates of the rest frame and
$n$ is an integer. This is an exact solution of
\begin{equation}
\nabla_{\mu}\nabla^{\mu} w -\frac{dV}{dw} - F = 0\,,
\end{equation}
as long as $F$ is given by
\begin{eqnarray}
F &=& -\frac{a^2\sin^2\theta + n^2}{\rho^2_{\rm BL}}\sin(t+r) \cos(n\theta) \\ 
&& + \frac{2r}{\rho^2_{\rm BL}} \cos(t+r) \cos (n\theta) \nonumber\\
&& - \frac{n\cos\theta}{\rho^2_{\rm BL}\sin\theta} \sin(t+r)\sin
(n\theta)\nonumber\\
&& - \frac{dV}{dw} \,.\nonumber
\end{eqnarray}
Both $w$ and $F$ are scalar quantities.
The evolution equation (\ref{Eq:WEgen2}) is modified according to
\begin{eqnarray}
\partial_t T &=& -\Big(\gamma^{ti}\partial_i T +
\partial_i(\gamma^{it}T)+ \partial_i (\gamma^{ij}d_j)
\label{Eq:WEmodif2}\\
&& +
\partial_t \gamma^{tt}T + \partial_t
\gamma^{tj}d_j -\sqrt{-g}\frac{dV}{d\Phi} \nonumber\\
&&- \sqrt{-g}F \Big)/\gamma^{tt},
\nonumber
\end{eqnarray}
where $g = \det (g_{\mu\nu})$.  On the axis of symmetry we use the
limits $\lim_{\theta\to 0}\frac{\sin n\theta}{\sin\theta} = n$ and
$\lim_{\theta\to \pi}\frac{\sin n\theta}{\sin\theta} = (-1)^{n+1}n$.
The result of our convergence test for $M=1$, $\beta = -0.85$,
$a=0.99$, and $n=2$, confirms that we have fourth order convergence on
each grid and is shown in Fig.~\ref{Fig:Qa99_b85}.  The simulation was
stopped when the inner spherical grid was about to touch the outer
spherical grid.  

Furthermore, Fig.~\ref{Fig:nodiss} illustrates that failing to add artificial
dissipation to the scheme can lead to numerical instabilities.

\section{Conclusion}
\label{Sec:conclusion}

The difficulties associated with the construction of a stable scheme for
the wave equation on a black hole background are also present in
simulations of fully nonlinear general relativistic systems.
Therefore, the ability to handle a scalar field propagating around a
boosted spinning black hole is one of the first necessary steps
towards the construction of a successful, long term stable binary
black hole collision numerical code from which gravitational wave
templates can be extracted.  In this work we showed that the
overlapping grids method is compatible with fourth order accuracy,
provided that sufficiently high order interpolation and artificial
dissipation are used.  We also noticed that simple schemes at times
work better than more complicated techniques based on conservation of
discrete quantities (see, for example, the treatment of the axis in
Appendix \ref{App:Fourth}).

In Appendix \ref{App:Pireduction} we observed that different
hyperbolic first order reductions, which have different levels of
hyperbolicity (symmetric or only strongly hyperbolic) inside (or near)
the excision region, have different stability properties.  This was
unexpected since in the continuum limit the treatment of the excision
region, if contained inside the black hole, should not affect the
solution outside and therefore should not introduce any
instabilities\footnote{Although high frequency modes can travel in the
wrong direction and therefore can escape from the black hole, the
expectation is that artificial dissipation will suppress these
modes.}.  We noticed that for the $T$-formulation the behavior of the
solution is highly sensitive to the discretization (changing the way
the dissipative operator is modified at the inner boundary, for
example, can have noticeable effects on the growth rate of the error).
It would be interesting to combine the $\Pi$- and $T$-formulations by
using the former in the inner spheroidal patch and the latter in the
other patches, closely resembling the ``interpolating formulation''
introduced in \cite{CLNPRST}.

In this work, as in \cite{CN}, we only considered fully first order
systems.  Recently, second order in space systems have generated more
interest, both at the continuum and discrete level
\cite{NOR,GG1,GG2,KO,KPY,MatNor,C1,Bel}.  We intend to apply the
overlapping grid method to such systems and, most importantly, to the
fully nonlinear dynamical case, in which non trivial issues arise.
Unlike in the toy model cases that we have investigated so far, in the
nonlinear case one has to face the problem of tracking a suitable
outflow surface containing the black hole singularity and the
dynamical generation of a grid adapted to this surface.  We are
currently working on a number of related points.

Overlapping grids represent possibly the simplest and most flexible
technique capable of accurately handling smooth and time dependent
boundaries within finite differencing.  It is our hope that it will
become a useful tool for the the binary black hole problem.


\begin{acknowledgments}

We would like to thank Eric Hirschmann for many interesting
discussions.  This research was supported in part by NSF and
NASA under grants
NSF-PHY-0244699, NSF-PHY-0326311 and NASA-NAG5-13430 
to Louisiana State University and the
Horace Hearne Jr. Labratory for Theoretical Physics.  GC was also
supported by a Marie Curie Intra-European Fellowship within the 6th
European Community Framework Program.

\end{acknowledgments}

\appendix 

\section{Modification of the 4th order accurate operator}
\label{App:43operator}

The fourth order accurate centered difference operator is given by
\[
Du_{i}  = (-u_{i+2} + 8u_{i+1} - 8u_{i-1} + u_{i-2})/(12h)
\]
When no boundaries are present (or if the gridfunctions are periodic),
we have
\begin{equation}
(u,Dv)_h = -(Du,v)_h\,,
\end{equation}
where $(u,v)_h = h\sum_j u_j v_j$.  In \cite{GKO-Book} it is shown
that, by appropriately modifying the difference operator and the
scalar product near and at the boundary, one can recover the summation
by parts rule \cite{Str}
\begin{equation}
(u,Dv)_h = -(Du,v)_h + u_jv_j|^N_0 . \label{Eq:SBP}
\end{equation}

We are interested in a globally fourth order accurate scheme.  For
this purpose one could use a sixth order accurate operator at the
interior with a third order accurate modification near the boundaries,
or a fourth order accurate operator in the interior with a third order
accurate modification near the boundaries.  Whereas the first operator
satisfies the summation by parts rule with respect to a diagonal
scalar product, the second one requires a non-diagonal scalar product
\begin{equation}
(u,v)_h = h \sum_{i,j=0}^N u_iv_j\sigma_{ij}\,.
\end{equation}
Since the global accuracy is the same, in this work we choose the
second option as it involves a smaller stencil and therefore requires
fewer computational resources.  Its modification near the boundary is
given by
\begin{eqnarray}
Du_i &=& \frac{1}{h} \sum_{j=0}^6 d_{ij} u_j  \qquad i = 0,\ldots,4\\
Du_i &=& \frac{1}{h} \sum_{j=0}^6 -d_{N-i,j} u_{N-j} \qquad i =
N-4,\ldots, N
\end{eqnarray}
where 
\begin{eqnarray*}
d_{00} &=& -11/6\\
d_{01} &=& 3\\
d_{02} &=& -3/2\\
d_{03} &=& 1/3\\
d_{04} &=& 0\\
d_{05} &=& 0\\
d_{06} &=& 0\\
d_{10} &=&
-24 (-779042810827742869+\\
&&104535124033147 \sqrt{26116897})/f_1\\
d_{11} &=&
-1/6 (-176530817412806109689+\\
&&29768274816875927 \sqrt{26116897})/f_1\\
d_{12} &=&
343 (-171079116122226871+\\
&&27975630462649 \sqrt{26116897})/f_1\\
d_{13} &=&
-3/2 (-7475554291248533227+\\
&&1648464218793925 \sqrt{26116897})/f_1\\
d_{14} &=&
1/3 (-2383792768180030915+\\
&&1179620587812973 \sqrt{26116897})/f_1\\
d_{15} &=& -1232 (-115724529581315+\\
&&37280576429 \sqrt{26116897})/f_1\\
d_{16} &=& 0\\
d_{20} &=& -12 (-380966843+86315 \sqrt{26116897})/f_2\\
d_{21} &=& 1/3 (5024933015+2010631 \sqrt{26116897})/f_2\\
d_{22} &=& -231/2 (-431968921+86711 \sqrt{26116897})/f_2\\
d_{23} &=& (-65931742559+12256337 \sqrt{26116897})/f_2\\
d_{24} &=& -1/6 (-50597298167+9716873 \sqrt{26116897})/f_2\\
d_{25} &=& -88 (-15453061+2911 \sqrt{26116897})/f_2\\
d_{26} &=& 0\\
d_{30} &=& 48 (-56020909845192541+\\
&&9790180507043 \sqrt{26116897})/f_1\\
d_{31} &=&
1/6 (-9918249049237586011+\\
&&1463702013196501 \sqrt{26116897})/f_1\\
d_{32} &=&
-13 (-4130451756851441723+\\
&&664278707201077 \sqrt{26116897})/f_1\\
d_{33} &=&
3/2 (-26937108467782666617+\\
&&5169063172799767 \sqrt{26116897})/f_1\\
d_{34} &=&
-1/3 (6548308508012371315+\\
&&3968886380989379 \sqrt{26116897})/f_1\\
d_{35} &=&
88 (-91337851897923397+\\
&&19696768305507 \sqrt{26116897})/f_1\\
d_{36} &=& 242 (-120683+15 \sqrt{26116897})/f_3\\
d_{40} &=& 264 (-120683+15 \sqrt{26116897})/f_3\\
d_{41} &=& 1/3 (-43118111+23357 \sqrt{26116897})/f_3\\
d_{42} &=& -47/2 (-28770085+2259 \sqrt{26116897})/f_3\\
d_{43} &=& -3 (1003619433+11777 \sqrt{26116897})/f_3\\
d_{44} &=& -11/6 (-384168269+65747 \sqrt{26116897})/f_3\\
d_{45} &=& 22 (87290207+10221 \sqrt{26116897})/f_3\\
d_{46} &=& -66 (3692405+419 \sqrt{26116897})/f_3
\end{eqnarray*}
and
\begin{eqnarray*}
f_1 &=& -56764003702447356523\\
&&+8154993476273221 \sqrt{26116897}\\
f_2 &=& -55804550303+9650225 \sqrt{26116897}\\
f_3 &=& 3262210757+271861 \sqrt{26116897}
\end{eqnarray*}

The expression for the $\sigma_{ij}$ coefficients can be found in
\cite{GKO-Book}, where stability proofs for linear hyperbolic problems
without corners can be found.

\section{Higher order accurate discretizations near the axis of symmetry}
\label{App:Fourth}

To illustrate the difficulties that arise with higher order
discretizations of axisymmetric systems we consider the polar wave
equation written in first order form
\begin{eqnarray}
\partial_t T &=& \frac{1}{\rho} \partial_\rho (\rho P)\\
\partial_t P &=& \partial_\rho T
\end{eqnarray}
where $P|_{\rho = 0} = 0$.
This system admits the following conserved energy
\begin{equation}
E = \int_0^{\rho_{\rm max}} (T^2+P^2) \rho d\rho
\end{equation}
in the sense that its time derivative gives only boundary
contributions
\begin{equation}
\frac{dE}{dt} = 2(\rho T P)|_{\rho_{\rm max}}\,.
\end{equation}
From the vanishing of $P$ at $\rho =0$ we have that 
\begin{equation}
\lim_{\rho \to 0} \frac{1}{\rho} \partial_\rho (\rho P) = 2
\partial_\rho P
\end{equation}
Hence we consider the following semidiscrete approximation
\begin{eqnarray}
\frac{d}{dt} T_0 &=& 2 (DP)_0 \label{Eq:polar1}\\
\frac{d}{dt} T_i &=& \frac{1}{\rho_i} (D \rho P)_i\qquad i=1,\ldots, N
\label{Eq:polar2}\\
\frac{d}{dt} P_i &=& (D T)_i\qquad i=1,\ldots, N\label{Eq:polar3}
\end{eqnarray}
and discrete energy
\begin{equation}
E = \sum_{i=1}^N (T^2_i + P^2_i) \rho_i \sigma_i h + \alpha T_0^2 h^2 \,.
\label{Eq:polarE}
\end{equation}
Let us assume that $D$ is a finite difference operator satisfying the
summation by parts rule (\ref{Eq:SBP}) with respect to a diagonal
scalar product.  The time derivative of (\ref{Eq:polarE}) gives
\begin{eqnarray*}
\frac{d}{dt} E &=& 2\sum_{i=1}^N \left(T_i D(\rho P)_i + P_i\rho_i
DT_i\right)\sigma_i h + 4\alpha T_0 DP_0 h^2\\
&=& 2 T_N\rho_N P_N + 2 T_0 \left( 2\alpha DP_0 h - \sigma_0 D (\rho
P)_0\right)h \,.
\end{eqnarray*}
In order to have an energy estimate consistent with the one of the continuum 
we need to ensure that the last term vanishes, i.e., the difference
operator at the axis has to satisfy
\begin{equation}
2\alpha DP_0 h = \sigma_0 D(\rho P)_0 \label{Eq:axiterm}
\end{equation}
for some $\alpha$.  As was pointed out in \cite{CN}, if a first order
accurate one sided difference operator is used on the axis, then
(\ref{Eq:axiterm}) is satisfied for $\alpha = \sigma_0/2 = 1/4$.

Using Maple we were able to obtain a (non unique) locally second order
accurate modification of the difference operator which leads to
discrete energy conservation.  However, we had to exploit the
regularity conditions at the axis ($P$ is odd and $T$ and $\rho P$ are
even functions of $\rho$).  The semidiscrete system
\begin{eqnarray}
\frac{d}{dt} T_0 &=& (3 P_1 + 4P_2-3P_3)/h \label{Eq:polcons}\\
\frac{d}{dt} T_1 &=& \frac{1}{\rho_1}((\rho P)_2 + 2 (\rho P)_3 )/(11h)\nonumber\\
\frac{d}{dt} T_2 &=& \frac{1}{\rho_2}(- 3(\rho P)_1 + 11(\rho P)_3
-2 (\rho P)_4)/(16h)\nonumber\\
\frac{d}{dt} T_3 &=& \frac{1}{\rho_3}(- 6(\rho P)_1 -11(\rho P)_2 \nonumber\\
&&+16(\rho P)_4 -2(\rho P)_5)/(26h)\nonumber\\
\frac{d}{dt} T_i &=& \frac{1}{\rho_i} D(\rho P)_i\qquad\qquad
i=4,\ldots,N-4\nonumber\\
\frac{d}{dt} T_{N-3} &=& (8(\rho P)_{N-5}-64(\rho P)_{N-4}+59(\rho
P)_{N-2}\nonumber\\
&&-3(\rho P)_N)/(98h)\nonumber\\
\frac{d}{dt} T_{N-2} &=& (8(\rho P)_{N-4}-59(\rho P)_{N-3}+59(\rho
P)_{N-1}\nonumber\\
&&-8(\rho P)_N)/(86h)\nonumber\\
\frac{d}{dt} T_{N-1} &=& ((\rho P)_{N}-(\rho P)_{N-2})/(2h)\nonumber\\
\frac{d}{dt} T_N &=& (3(\rho P)_{N-3}+8(\rho P)_{N-2}-59(\rho
P)_{N-1}\nonumber\\
&&+48(\rho P)_N)/(34h)\nonumber \\
\frac{d}{dt} P_1 &=& (-3 T_0 + T_2 + 2 T_3 )/(11h)\nonumber\\
\frac{d}{dt} P_2 &=& (-6 T_0 - 3T_1 + 11T_3
-2 T_4)/(16h)\nonumber\\
\frac{d}{dt} P_3 &=& (3T_0 - 6T_1 -11T_2 +
16T_4 -2T_5)/(26h)\nonumber\\
\frac{d}{dt} P_i &=& DT_i\qquad\qquad i=4,\ldots,N-4\nonumber\\
\frac{d}{dt} P_{N-3} &=& (8T_{N-5}-64T_{N-4}+59T_{N-2}-3T_N)/(98h)\nonumber\\
\frac{d}{dt} P_{N-2} &=& (8T_{N-4}-59T_{N-3}+59T_{N-1}-8T_N)/(86h)\nonumber\\
\frac{d}{dt} P_{N-1} &=& (T_{N}-T_{N-2})/(2h)\nonumber\\
\frac{d}{dt} P_N &=& (3T_{N-3}+8T_{N-2}-59T_{N-1}+48T_N)/(34h)\nonumber \\
\end{eqnarray}
where $D$ is the standard fourth order accurate centered approximation
of $\p_\rho$, conserves the following discrete energy
\begin{equation}
E = \sum_{i=1}^{N} (T_i^2 + P_i^2) \rho_i\sigma_ih +
T_0^2\sigma_0h^2\,,
\end{equation}
where $\sigma = \{ \frac{1}{8},\frac{11}{8},\frac{2}{3},\frac{13}{12},
1, \ldots , 1, \frac{49}{48}, \frac{43}{48}, \frac{59}{48},
\frac{17}{48} \}$.  We have
\begin{equation}
\frac{d}{dt} E = 2 T_N\rho_N P_N \,.
\end{equation}
Numerical experiments done at high resolutions suggest that the
overall order of accuracy of the scheme is three.

We now compare this discretization with a fourth order accurate
approximation, obtained by simply
exploiting the regularity conditions of the fields.  We use the
approximation (\ref{Eq:polar1}--\ref{Eq:polar3}) where $D$ is the
standard centered fourth order accurate finite difference operator,
appropriately modified at the outer boundary and use the fact that $T$
and $\rho P$ are symmetric across the axis.  We see that the time
derivative of the discrete energy (\ref{Eq:polarE}) is
\begin{eqnarray*}
\frac{d}{dt} E &=& 2T_N\rho_NP_N - \frac{1}{6}
(8T_0P_1-2T_0P_2-2T_1P_1)h \\
&&+\frac{4\alpha}{6}T_0(-P_2+8P_1)h
\end{eqnarray*}
Clearly, there is no value of $\alpha$ that leads to cancellations of
the terms near the axis.  Furthermore, if corrections of the form
$T_i^2h^2$ or $P_i^2h^2$ are added to the energy, products like
$T_iP_{i+2}$ or $P_iT_{i+2}$ appear in the estimate.  Also, any mixed
product $P_iT_jh^2$ would give rise to $P^2$ and $T^2$ terms.  This
discretization does not conserve the energy (\ref{Eq:polarE}).  In
particular for $\alpha=1/4$ one is left with the axis terms
\begin{equation}
\frac{1}{6}(T_0P_2+2T_1P_1)h\,.
\label{Eq:leftover}
\end{equation}

\begin{figure}[t]
\begin{center}
\includegraphics*[width=8cm]{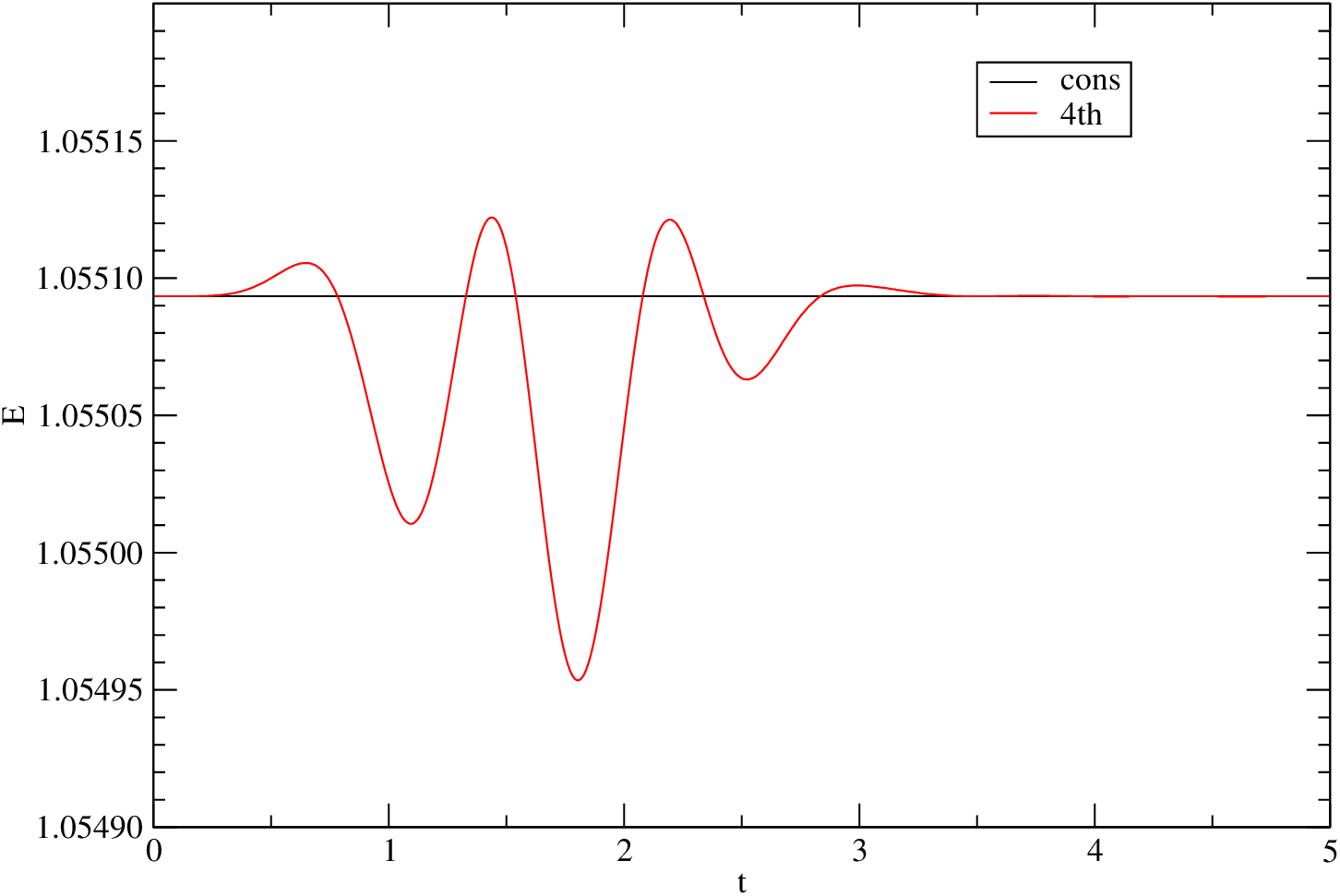}
\caption{We compare the energy conserving discretization
  (\ref{Eq:polcons}) with the simpler fourth order accurate one based
  on regularity.  We plot the discrete energy (\ref{Eq:polarE}) with
  $\alpha = 1/4$ as a function of time.  The grid consist of $200$
  gridpoints and the domain is $0 \le \rho \le 20$.  We use 4th order
  Runge-Kutta with a small Courant factor, $\Delta t/ \Delta \rho =
  1/10$ to be close to the semidiscrete approximation.  We use initial
  data $T = \sin^6(\rho)$ for $0\le \rho\le \pi$ and zero elsewhere.
  The energy of the simpler approximation oscillates about the value
  of the energy of the energy conserving scheme.}
\label{Fig:cons_vs_4th}
\includegraphics*[width=8cm]{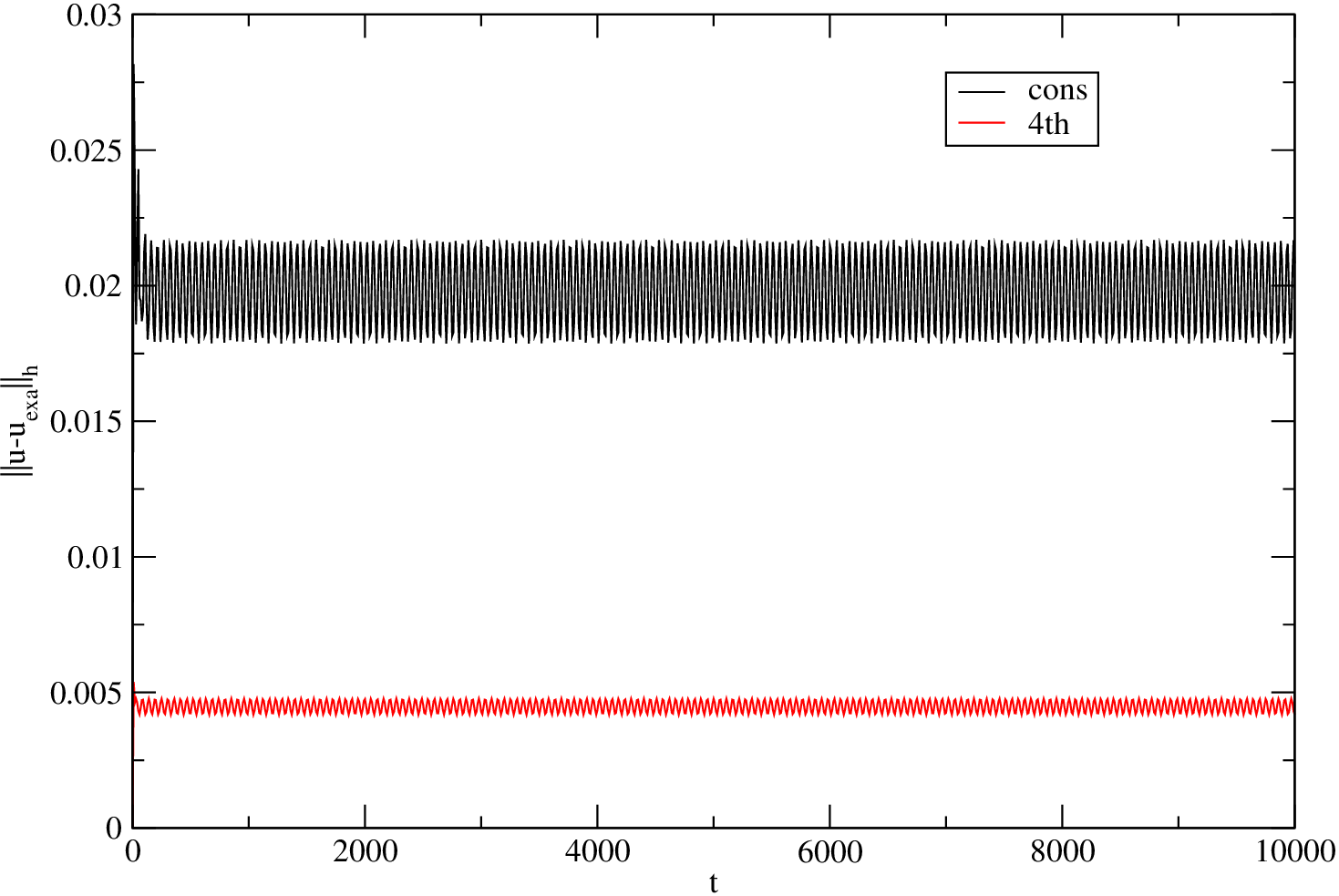}
\caption{We compare the numerical errors of the energy conserving
  discretization with the fourth order accurate one.  We use $50$
  gridpoints and the domain $0 \le \rho \le 5$.  The Courant factor is
  $\Delta t/ \Delta \rho = 1$.  We use the forcing solution method
  with exact solution $\Phi = \rho \sin\rho \sin t$ and forcing term
  $F = -\sin t (\sin\rho+3\rho \cos\rho)/\rho$.  The errors of the
  simpler fourth order accurate approximation are clearly much smaller
  than those of the energy conserving scheme.}
\label{Fig:smaller_error}
\end{center}
\end{figure}

Clearly, the fourth order accurate discretization based on regularity
conditions, although not energy conserving, is much simpler.  The
introduction of ghost zones makes it numerical implementation trivial.
Fig.~\ref{Fig:cons_vs_4th} illustrates how, for a particular problem
and with some particular initial data, the energy of the fourth order
accurate operator oscillates about the value of the energy conserving
scheme.

Furthermore, no long term growth of the error was observed with the
simpler fourth order approximation.  In fact, our experiments indicate
that, in general, the errors are much smaller than with the energy
conserving scheme (\ref{Eq:polcons}).  This is due to the lower order
of local accuracy near the axis of symmetry and is illustrated in
Fig.~\ref{Fig:smaller_error}.

In our overlapping grid code we adopted the simpler fourth order
accurate approximation near the axis.

\section{A comparison between two first order reductions}
\label{App:Pireduction}

In \cite{CN} we performed long term stability tests to check that the
interpolation process did not introduce unwanted growth of the error.
There we used $M=a=\beta=0$ and had only the spherical inner patch and
the cylindrical patch.  Here we repeat the same test, but with
nontrivial values for the black hole mass and spin.  Interestingly,
this test revealed that the approximation for the initial-boundary
value problem for the reduction (\ref{Eq:WEgen1})--(\ref{Eq:WEgen3}),
which we will refer to as $T$-reduction from now on, suffers from
exponential growth of the error whenever the domain contains a region
in which the system is only strongly hyperbolic.  These errors are
concentrated near the excision region and, at the resolutions that we
typically use, become evident only after about a hundred crossing times.

To exclude the possibility of the overlapping grid method being
responsible for this growth, we eliminated all grids, except the inner
one.  The growth was still there.  Furthermore, the growth was present
also when the second order accurate, energy conserving method for
axisymmetric systems developed in \cite{CN} and the modification of
the dissipative operator constructed in \cite{LongPaper} were used.

We recall that the first order formulation
(\ref{Eq:WEgen1})--(\ref{Eq:WEgen3}) is symmetric hyperbolic only
outside the ergoregion (outside the black hole in the non rotating
case).  It is important to realize that the use of the discrete energy
method to obtain stable discretization relies on the fact that the
energy is a positive definite quadratic form of the main variables.
Not having symmetric hyperbolicity on the entire domain, we have no
guarantee of obtaining a stable discretization.

This prompted us to investigate a different first order reduction,
which, unlike the $T$-reduction, is symmetric hyperbolic everywhere.
The energy associated with the symmetrizer, however, is not conserved
and therefore we have no preferred way of discretizing the system.
After describing this alternative reduction we experimentally compare
the stability properties of the two formulations.

As in Sec.~\ref{Sec:wave_equation} we start with the wave equation
around a curved background
\[
\frac{1}{\sqrt{-g}} \p_{\mu} (\sqrt{-g} g^{\mu\nu}\p_{\nu} \Phi)
-\frac{dV}{d\Phi} = 0
\]
Using the ADM decomposition of the metric \cite{ADM} 
\begin{equation}
ds^2 = g_{\mu\nu}dx^{\mu}dx^{\nu} = -\alpha^2 dt^2 + h_{ij}(dx^i + \beta^i dt)(dx^j + \beta^j dt) \label{Eq:ADMmetric}
\end{equation}
the wave equation becomes
\begin{eqnarray*}
&&\p_t \left( \frac{\sqrt{h}}{\alpha} \left( \p_t \Phi - \beta^i\p_i
  \Phi\right) \right) =\\
&& \p_i \left( \frac{\sqrt{h}}{\alpha} \left(
  \beta^i \p_t \Phi + ( \alpha^2 h^{ij} - \beta^i \beta^j) \p_j
  \Phi\right) \right)-\alpha\sqrt{h}\frac{dV}{d\Phi} 
\end{eqnarray*}
where $h = \det(h_{ij})$.

We introduce the auxiliary variables $\Pi = (\beta^i \partial_i \Phi -
\partial_t \Phi)/\alpha$ and $d_i = \partial_i \Phi$.  Defining
\begin{eqnarray*}
\alpha K &=& \frac{1}{\sqrt{h}} \p_i (\sqrt{h} \beta^i) - \p_t
\ln\sqrt{h}\\
h^{ij}\Gamma^{k}_{ij} &=& -\frac{1}{\sqrt{h}}\p_i (\sqrt{h}
h^{ki})
\end{eqnarray*}
the wave equation can be written in the form
\begin{eqnarray}
\frac{\partial \Phi}{\partial t} &=& \beta^i \partial_i \Phi - \alpha \Pi,\\
\frac{\partial \Pi}{\partial t} &=& \beta^i \partial_i \Pi  
       - \alpha h^{ij} \partial_j d_i
       -\alpha h^{ij}\Gamma^{k}{}_{ij}d_k \nonumber\\
      &&- h^{ij}\partial_i \alpha d_j
       + \alpha K \Pi + \alpha\frac{dV}{d\Phi} \\
\frac{\partial d_i}{\partial t} &=&  \beta^j\partial_j d_i 
       - \alpha\partial_i \Pi
       + \partial_i \beta^j d_j  - \partial_i \alpha \Pi
\label{Eq:Pireduction}
\end{eqnarray}
We will refer to this first order reduction of the wave equation as
the $\Pi$-reduction.  In the particular case of the Kerr-Schild metric
we have
\begin{eqnarray*}
H &=& \frac{Mr}{\rho^2_{\rm BL}}\\
\alpha &=& (1+2H)^{-1/2}\\
\beta^r &=& 2H\alpha^2 = 1-\alpha^2\\
\beta^\theta &=& 0\\
\sqrt{h} &\equiv& \det(h_{ij})^{1/2} = \alpha^{-1} \rho^2_{\rm BL} \sin\theta\\
h^{rr} &=& \alpha^2 + \frac{a^2\sin^2\theta}{\rho^2_{BL}}\\
h^{r\theta} &=& 0\\
h^{\theta\theta} &=&  \frac{1}{\rho^2_{\rm BL}}
\end{eqnarray*}

The principal part is given by
\[
A^n = \left( \begin{array}{ccc}
\beta^n & 0 & 0 \\
0 & \beta^n & -\alpha h^{jn} \\
0 & -\alpha n_i & \beta^n \delta^{ij}
\end{array}\right)
\]
where $\beta^n = \beta^i n_i$ and $h^{jn} = h^{ji}n_i$.
The characteristic variables and speeds in the direction $n$ are 
\[
\begin{array}{r@{\,=\,}l@{\qquad}c}
w^{(\pm;n)} & \Pi \pm \sqrt{h^{nn}} d_n\,, & \beta^n \mp \alpha \sqrt{h^{nn}}\\
w^{(0;n)}_A & d_A\,, & \beta^n\\
w^{(0;n)}_\Phi & \Phi\,, & \beta^n 
\end{array}
\]
where $A$ is a vector orthogonal to $n$.  The inverse transformation
is
\begin{eqnarray*}
\Pi &=& \frac{1}{2}(w^{(+;n)} + w^{(-;n)})\\
d_n &=& \frac{1}{2\sqrt{h^{nn}}}(w^{(+;n)} - w^{(-;n)})\\
d_A &=& w^{(0;n)}_A\\
\Phi &=& w^{(0;n)}_\Phi
\end{eqnarray*}

The most general symmetrizer is
\[
H(t,\vec{x}) = \left(
\begin{array}{ccc}
\xi & 0 & 0 \\
0 & \eta & 0 \\
0 & 0 & \eta h^{ij}
\end{array}
\right)
\]
where $\xi=\xi(t,\vec x) > 0$ and $\eta = \eta(t,\vec x) > 0$.  This
reduction, unlike the $T$-reduction, has an everywhere positive
definite symmetrizer (it is symmetric hyperbolic) if and only if the
$t={\rm const.}$ slices of spacetime are space-like.

Using the regularity conditions (the fields $\Phi$, $\Pi$, $d_r$ are
even and $d_{\theta}$ is odd) and the fact that on the axis
$\p_{\theta} \alpha = 0$ and $\alpha h^{ij} \Gamma^{\theta}_{ij}
d_{\theta} = - \alpha h^{\theta\theta} \p_{\theta} d_{\theta}$ we
obtain the following regularized equation on the axis
\begin{eqnarray*}
\p_t \Pi &=& \beta^r \p_r \Pi - \alpha h^{rr} \p_r d_r - 2\alpha
h^{\theta\theta} \p_{\theta} d_{\theta} + \alpha h^{ij}
\Gamma^{r}_{ij} d_r \\
&&-h^{rr} \p_r \alpha d_r + \alpha K \Pi + \alpha \frac{dV}{d\Phi}
\end{eqnarray*}

We compare single grid, long term stability of the $T$- and
$\Pi$-formulations using the same second order accurate finite
difference operator modified as in \cite{CN}, where one sided
difference operators were used at the physical boundaries and at the
axis.  The dissipative operator is modified according to prescription
given in \cite{LongPaper} at the physical boundaries and regularity is
used at the axis.  No special grouping of variables is done for the
$\Pi$-formulation.  Long term stability tests were done using the
forcing solution of Sec.~\ref{Sec:NumExp} with $n=2$.  At the outer
boundary $r=r_{\rm max}$ we give data to the incoming characteristic
variables of $A^n$. (In the $T$-formulation we actually give data to
the incoming characteristic variables of $HA^n$.  One can show that
both methods control the boundary term arising in the energy estimate
\cite{CalPhD}.)

\begin{figure}[t]
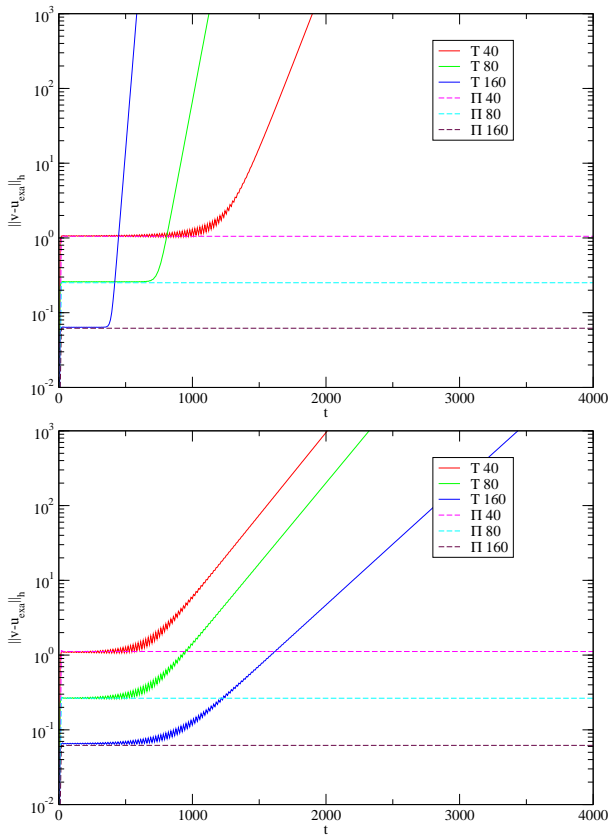

\begin{center}
\includegraphics*[width=8cm]{longsigma0.02.eps}
\includegraphics*[width=8cm]{longsigma0.02_a0.99.eps}
\caption{Long term stability test comparing two different reductions
  of the wave equation ($T$ and $\Pi$) at different resolutions
  using the second order accurate discretization.  We used the
  following parameters: $r_{\rm min} = 1$, $r_{\rm max} = 10$, $t_{\rm
  max} = 5\cdot 10^4$, $k/h = 1.5$, $\sigma = 0.02$, $M = 1$, $a=0$
  (top) or $a=0.99$ (bottom), and $n=2$.}
\label{Fig:longsigma0.02}
\end{center}
\end{figure}

Our tests revealed that the $\Pi$-formulation, despite not being
energy conserving, outperformes the $T$-formulation in terms of
stability and long term stability.  Fig.~(\ref{Fig:longsigma0.02})
clearly shows that if the excision boundary is placed well inside the
event horizon the $T$-discretization is unstable, whereas the
$\Pi$-discretization is stable.  Interestingly, in the highly spinning
case, the rate of growth of the error in the $T$-formulation does not
seem to increase with resolution.  Note that by excising at $r_{\rm
min} = M = 1$, the region of lack of symmetric hyperbolicity is
actually larger in the non-spinning case.

In an attempt to understand the different behavior of the $T$- and
$\Pi$-reductions of the wave equation, we perform a Laplace-Fourier
analysis of an analogous constant coefficient toy model
initial-boundary value problem.

Consider the wave equation around Minkowski in Cartesian coordinates
\begin{equation}
\eta^{\mu'\nu'} \p_{\mu'}\p_{\nu'} \phi = 0
\end{equation}
where $\eta^{\mu'\nu'} = {\rm diag} \{ -1,+1,+1,+1\}$.  Under the
following change of coordinates
\begin{equation}
t = t'\,,\quad x^i = x'^i - \beta^i t'
\end{equation}
where $\beta^i$ are constant, the wave equation takes the form
\begin{equation}
\p_t^2 \phi = 2 \beta^i \p_t\p_i \phi + (\delta^{ij} - \beta^i\beta^j)
\p_i\p_j \phi
\end{equation}
We consider two different types of reductions: the $T$-reduction
\begin{eqnarray}
\p_t \phi &=& T \\
\p_t T &=& 2\beta^i\p_i T + (\delta^{ij} - \beta^i\beta^j) \p_id_j\\
\p_t d_j &=& \p_j T
\end{eqnarray}
and the $\Pi$-reduction 
\begin{eqnarray}
\p_t \phi &=& \beta^i \p_i \phi + \Pi\\
\p_t \Pi &=& \beta^i \p_i \Pi + \delta^{ij} \p_i d_j\\
\p_t d_j &=& \beta^i \p_i d_j + \p_j \Pi
\end{eqnarray}
Notice that the constraints $C_j \equiv d_j - \p_j \phi = 0$, which
are introduced in the reduction process, propagate differently in the
two formulations.  Whereas in the first case we have $\p_t C_j = 0$,
in the second case we have $\p_t C_j = \beta^i \p_i C_j$.  In both
cases the $\phi$ variable decouples from the system. We will drop it
in the analysis that follows.

The $T$-reduction has the following symmetrizer
\begin{equation}
T^2 + (\delta^{ij} - \beta^i \beta^j) d_id_j \label{Eq:symmetrizerT}
\end{equation}
and characteristic speeds
\begin{equation}
0, \, \beta^n \pm 1\,.
\end{equation}
The $\Pi$-reduction has a simpler symmetrizer
\begin{equation}
\Pi^2 + \delta^{ij}d_id_j \label{Eq:symmetrizerPi}
\end{equation}
and the characteristic speeds are given by 
\begin{equation}
\beta^n,\, \beta^n \pm 1\,.
\end{equation}
Most importantly, whereas (\ref{Eq:symmetrizerPi}) is positive
definite for any $\beta^i$, (\ref{Eq:symmetrizerT}) is positive
definite if and only if $\delta_{ij}\beta^i\beta^j < 1$.  The last
condition is equivalent to the requirement that the
vector field $\p_t$ be time-like.

Assume $\beta_1 > 1$ and consider the quarter space problem $x>0$ with
periodic solutions in $y$ and $z$ and no boundary conditions at $x=0$.
With the $\Pi$-reduction energy estimates can be obtained in the
outflow case using the energy method.  For the $T$-formulation a
Laplace-Fourier analysis yields the following eigenvalue problem (a
system of ordinary differential equations in the variable $x$)
\begin{eqnarray*}
\p_x T &=& sd_1\\
(\beta_1^2-1) \p_x d_1 &=& 2\beta_1 (s-i\beta_A\omega^A) d_1
\\
&& -s^{-1}[(s-i\beta_A\omega^A)^2 + \omega_A\omega^A] T
\end{eqnarray*}
If an eigenvalue with $\Re(s)>0$ exists, then the initial-boundary
value problem is ill-posed in any sense \cite{GKO-Book}.  The
eigenvalues are
\begin{equation}
\lambda_{\pm} = \beta_1 (s-i\beta_A\omega^A) \pm \sqrt{
  (s-i\beta_A\omega^A)^2 - (\beta_1^2-1)\omega_A\omega^A} \label{Eq:eigLF}
\end{equation}
Because of the requirement that the solution belongs to
$L_2(0,+\infty)$, we must discard those eigenvalues which have
positive real part.  Since $\Re (\lambda_{\pm}) >0$ if $\Re(s) >0$,
the problem is not obviously ill-posed.

This analysis reinforces our suspicion that the observed exponential
frequency dependent growth is merely due to the discretization.  More
work is needed to exactly establish the cause of the growth.  It is
possible that applying the Laplace transform method to the
semi-discrete problem may shed some light.



\begin{thebibliography}{99}

\bibitem{Unr}
W.~Unruh, quoted in J. Thornburg, Class. Quantum Grav. {\bf 4}, 1119 (1987).


\bibitem{L}
L.~Lehner, Class.~Quantum Grav.~{\bf 18}, R25 (2001).

\bibitem{BS}
T.W.~Baumgarte and S.L.~Shapiro, Phys.~Rept.~{\bf 376} 41-131 (2003).

\bibitem{S}
B.F.~Schutz, gr-qc/0410121.

\bibitem{A}
M.~Alcubierre, gr-qc/0412019.

\bibitem{BTJ} B.~Br\"ugmann, W.~Tichy and N.~Jansen, 
       Phys.~Rev.~Lett.~{\bf 92}, 211101 (2004).

\bibitem{ABDGHHH} M.~Alcubierre, B.~Br\"ugmann, P.~Diener, F.S.~Guzm\'an,
   I.~Hawke, S.~Hawley, F.~Herrmann, M.~Kopptiz, D.~Pollney, E.~Seidel
   and J.~Thornburg, gr-qc/0411149 (2004).

\bibitem{ABDHPST} M.~Alcubierre, B.~Br\"ugmann, P.~Diener,
   F.~Herrmann, D.~Pollney, E.~Seidel
   and R.~Takahashi, gr-qc/0411137 (2004).


\bibitem{BBHGCA}
Binary Black Hole Grand Challenge Alliance, Phys.~Rev.~Lett.~{\bf 80},
     2512 (1998).

\bibitem{Brandt}
S.~Brandt, R.~Correll, R.~G\' omez, M.~Huq, P.~Laguna, L.~Lehner, 
  P.~Marronetti, R.A.~Matzner, D.~Neilsen, J.~Pullin, E.~Schnetter, 
  D.~Shoemaker, and J.~Winicour, Phys.~Rev.~Lett.~{\bf 85}, 5496 (2000).

\bibitem{YoBauSha}
H.~Yo, T.W.~Baumgarte, and S.L.~Shapiro, Phys.~Rev.~D {\bf 64}, 124011
(2001).

\bibitem{ShoSmiSpeLagSchFis} 
D.~Shoemaker, K.L.~Smith, U.~Sperhake,
P.~Laguna, E.~Schnetter, and D.~Fiske, Class.~Quant.~Grav.~{\bf 20}
3729-3744 (2003).

\bibitem{SpeSmiKelLagSho}
U.~Sperhake,  K.L.~Smith, B.~Kelly, P.~Laguna, and D.~Shoemaker,
Phys.~Rev.~{\bf D} 69, 024012 (2004).

\bibitem{CN}
G.~Calabrese and D.~Neilsen, Phys.~Rev.~D {\bf 69}, 044020 (2004).

\bibitem{Tho_excision}
J.~Thornburg, Class.~Quantum Grav.~{\bf 21} 3665-3691 (2004).

\bibitem{Tho_AH}
J.~Thornburg, Class.~Quantum Grav.~{\bf 21} 743-766 (2004).

\bibitem{And}
M.W.~Anderson, {\it Constrained evolution in numerical relativity.}
Ph.D. thesis, University of Texas at Austin, 2002.

\bibitem{ReuTigLeh}
O.~Reula, M.~Tiglio, L.~Lehner, in preparation.

\bibitem{SAT}
M.H.~Carpenter, D.~Gottlieb, and S.~Abarbanel, J.~Comp.~Phys.~{\bf
  111} 220-236 (1994).

\bibitem{Sta}
G.~Starius, Numer.~Math.~{\bf 35}, 241-255 (1980).

\bibitem{GKO-Book}
B.~Gustafsson, H.~Kreiss, and J.~Oliger,
{\em Time dependent problems and difference methods}
(John Wiley \& Sons, New York, 1995).

\bibitem{OlsPet}
F.~Olsson and N.A.~Petersson, Computers and Fluids {\bf 25}, 583 (1996).

\bibitem{KL-Book} H.O.~Kreiss, J.~Lorenz, {\em Initial-Boundary
Value Problems and the Navier-Stokes Equations} (Academic Press, Boston, 1989).

\bibitem{KerSch}
R.P.~Kerr and A.~Schild in {\it Comitato Nazionale per le Manifestazioni
    Celebrative del IV Centenario della Nascita di Galileo Galilei, Atti
    del Convegno sulla Relativit\`a Generale:  Problemi dell'Energia
    e Onde Gravitazionali,} 1--12, edited by G.~Barb\' era, Florence, (1965);
R.P.~Kerr and A.~Schild, in {\it Proceedings of Symposia in Applied
    Mathematics} {\bf 17}, 199, American Math.~Soc.~(1965).

\bibitem{HE}
S.W.~Hawking and G.F.R.~Ellis, {\it The large scale structure of
space-time} (Cambridge University Press, Cambridge, 1973).

\bibitem{BoyLin}
R.H.~Boyer and R.W.~Lindquist, J.~Math.~Phys.~{\bf 8}, 265 (1967).

\bibitem{H}
R.~L.~Higdon, SIAM Review {\bf 28}, 2, 177-217 (1986).

\bibitem{GR}
E.~Godlewski and P.-A.~Raviart, {\em Numerical Approximations of
  Hyperbolic Systems of Conservation Laws} (Springer, New York, 1996).

\bibitem{R2}
O.~Reula, ``Strongly hyperbolic systems in General
Relativity'', gr-qc/0403007.

\bibitem{Ols}
P.~Olsson, Math.~Comp.~{\bf 64}, 1035 (1995); {\bf 64}, S23 (1995);
{\bf 64}, 1473 (1995).

\bibitem{Gus}
B.~Gustafsson, SIAM J.~Num.~Anal.~{\bf 18} 179-190 (1981).

\bibitem{CFL}
R.~Courant, K.O.~Friedrichs, and H.~Lewy, Math.~Ann.~{\bf 100}, 32 (1928).

\bibitem{CLNPRST}
G.~Calabrese, L.~Lehner, D.~Neilsen, J.~Pullin, O.~Reula, O.~Sarbach,
and M.~Tiglio, Class.~Quant.~Grav.~{\bf 20}, L245-L252 (2003).

\bibitem{NOR}
G.~Nagy, O.~Ortiz, and O.~Reula, Phys.~Rev.~D {\bf 70}, 044012 (2004).

\bibitem{GG1}
C.~Gundlach, J.M.~Martin-Garcia, Phys.~Rev.~D {\bf 70}, 044031 (2004).

\bibitem{GG2} 
C.~Gundlach, J.M.~Martin-Garcia, Phys.~Rev.~D {\bf 70}, 044032 (2004).

\bibitem{KO}
H.~Kreiss and O.~Ortiz, Lect.~Notes Phys.~{\bf 604}, 359 (2002).

\bibitem{KPY}
H.~Kreiss, N.~Petersson, and J.~Ystr\"om, SIAM J.~Numer.~Anal.~{\bf
  40}, 1940-1967 (2002); H.~Kreiss, N.~Petersson, and J.~Ystr\"om,
{\em Difference approximations of the Neumann problem for the second
  order wave equation}, UCRL-JC-153184, (2003). 

\bibitem{MatNor}
K.~Mattsson and J.~Nordstr\"om, J.~Comput.~Phys.~{\bf 199}, 503-540 (2004).

\bibitem{C1}
G.~Calabrese, Class.~Quant.~Grav.~{\bf 21}, 4025-4040 (2004).

\bibitem{C2}
G.~Calabrese, ``Finite differencing second order systems describing
black hole spacetimes'', gr-qc/0410062.

\bibitem{Bel} 
B.~Szilagyi, ``Summation by Parts in Numerical Relativity -- strengths
and limitations'', Talk at Sources and Simulations Seminar,
Pennsylvania State University, November 11, 2004.  Available at
http://cgpg.gravity.psu.edu/events/sss/2004/fall.shtml

\bibitem{Str}
B.~Strand, J.~Comput.~Phys.~{\bf 110}, 47 (1994).


\bibitem{LongPaper}
G.~Calabrese, L.~Lehner, O.~Reula, O.~Sarbach, and M.~Tiglio, 
  Class.~Quantum Grav.~{\bf 21} 5735-5757 (2004).

\bibitem{ADM}
R.~Arnowitt, S.~Deser, and C.~Misner, in {\it Gravitation: An Introduction
    to Current Research,} edited by L.~Witten (Wiley, New York, 1962).

\bibitem{CalPhD}
G.~Calabrese, {\it Constraint Preserving Boundary Conditions for the
Linearized Einstein Equations,} Ph.D. thesis, Louisiana State University,
etd-1105103-100340 (2003).


\end{thebibliography}
\end{document}